\documentclass[11pt,a4paper]{article}
\pdfoutput=1

\usepackage{graphicx}
\expandafter\let\csname equation*\endcsname\relax
\expandafter\let\csname endequation*\endcsname\relax

\usepackage[cp1251]{inputenc}
\usepackage{amsmath}
\usepackage{amssymb}
\usepackage{bm}
\usepackage{cancel}
\usepackage[normalem]{ulem}
\usepackage{subfig}
\usepackage{placeins}
\usepackage{color}
\usepackage{hyperref}

\begin{document}

\title{Two dimensional soliton in tumor induced angiogenesis}
\author{ L L Bonilla$^*$, M Carretero and F Terragni  \\
Gregorio Mill\'an Institute,\\
 Fluid Dynamics, Nanoscience and Industrial Mathematics,\\ 
 and Department of Mathematics,\\ 
 Universidad Carlos III de Madrid, 28911 Legan\'es, Spain\\
$^{*}$Corresponding author. E-mail: bonilla@ing.uc3m.es }

\date{\today}
\maketitle

\begin{abstract} 
Ensemble averages of a stochastic model show that, after a formation stage, the tips of active blood vessels in an angiogenic network form a moving two dimensional stable diffusive soliton, which advances toward sources of growth factor. Here we use methods of multiple scales to find the diffusive soliton as a solution of a deterministic equation for the mean density of active endothelial cells tips. We characterize the diffusive soliton shape in a general geometry, and find that its vector velocity and the trajectory of its center of mass along curvilinear coordinates solve appropriate collective coordinate equations. The vessel tip density predicted by the soliton compares well with that obtained by ensemble averages of simulations of the stochastic model. 
\end{abstract}



\section{Introduction}
\label{sec:intro}
Angiogenesis is the growth of blood vessels out of a primary vessel, a complex multiscale process that determines organ growth and regeneration, tissue repair, wound healing and many other natural operations \cite{car05,CT05,GG05,fru07,fig08,car11,ton00}. Its imbalance contributes to numerous malignant, inflammatory, ischaemic, infectious, and immune diseases \cite{car05}, such as cancer \cite{fol71,fol74,lio77,fol06,lia07,zua18}, rheumatoid arthritis \cite{mar06}, neovascular age-related macular degeneration \cite{jag08,niv14}, endometriosis \cite{tay09}, and diabetes \cite{mar03}. 

Normal angiogenesis proceeds as follows. During inflammation or under hypoxia, cells may activate signaling pathways that lead to secretion of pro-angiogenic proteins, such as Vessel Endothelial Growth Factor (VEGF). VEGF diffuses in the tissue, binds to extracellular matrix (ECM) components and forms a spatial concentration gradient in the direction of hypoxia. When VEGF molecules reach an existing blood vessel, they promote diminishing adhesion of its cells and the growth of new vessel sprouts. VEGF also activates the tip cell phenotype in endothelial cells (ECs) of the vessel, which then grow filopodia with many VEGF receptors. The tip cells pull the other ECs, open a pathway in the ECM, lead the new sprouts, and migrate in the direction of increasing VEGF concentration \cite{ger03}. {\em Branching of new sprouts} occurs as a result of signaling and mechanical cues between neighboring ECs \cite{hel07,jol15,veg20}. ECs in growing sprouts alter their shape to form a lumen connected to the initial vessel that is capable of carrying blood \cite{geb16}. Sprouts meet and merge in a process called {\em anastomosis} to improve blood circulation inside the new vessels. Poorly perfused vessels may become thinner and their ECs, in a process that inverts angiogenesis, may retract to neighboring vessels leading to a more robust blood circulation \cite{fra15}. Thus, the vascular plexus remodels into a highly organized and hierarchical network of larger vessels ramifying into smaller ones \cite{szy18}. In normal processes of wound healing or organ growth, the cells inhibit the production of growth factors when the process is finished.

The previous picture changes in significant ways in pathological angiogenesis. Hypoxic cells of an incipient tumor experience lack of oxygen and nutrients. Then they produce VEGF that induces angiogenesis, and new vessel sprouts exit from a nearby primary vessel and move in the tumor direction \cite{fol74,car05,fol06}. Blood brings oxygen and nutrients that foster tumor growth. Instead of inhibiting production of growth factors, tumor cells continue secreting growth factors that attract more vessel sprouts and facilitate tumor expansion. Together with experiments, there are many models spanning from the cellular to macroscopic scales that try to understand angiogenesis \cite{lio77,veg20,sto91,sto91b,cha93,cha95,orm97,and108,ton01,lev01,pla03,man04,sun05a,sun05,cha06,ste06,bau07,cap09,owe09,jac10,das10,swa11,sci11,sci13,cot14,dej14,ben14,bon14,hec15,per17,pil17,bon19}. 

Early models consider reaction-diffusion equations for densities of cells and chemicals (growth factors, fibronectin, etc.) \cite{lio77,cha93,cha95}. They cannot treat the growth and evolution of individual blood vessels. Tip cell stochastic models of tumor induced angiogenesis are among the simplest ones for this complex multiscale process. Their basic assumptions are that (i) the cells of a blood sprout tip do not proliferate and move towards the tumor producing growth factor, and (ii) proliferating stalk cells build the sprout along the trajectory of the sprout tip. Thus tip cell models are based on the motion of single particles representing the tip cells and their trajectories constitute the advancing blood vessels network \cite{sto91,sto91b,and108,pla03,man04,cap09,bon14,hec15,bon19,ter16}. Tip cell models describe angiogenesis over distances that are large compared with a cell size, thereby renouncing to detailed descriptions of cellular processes. More complex models include tip and stalk cell dynamics, the motion of tip and stalk cells on the extracellular matrix outside blood vessels, signaling pathways and EC phenotype selection, blood circulation in newly formed vessels, and so on \cite{bau07,jac10,ben14,hec15,ber18,veg20}. 

\begin{figure}[h]
\begin{center}
\includegraphics[width=14cm]{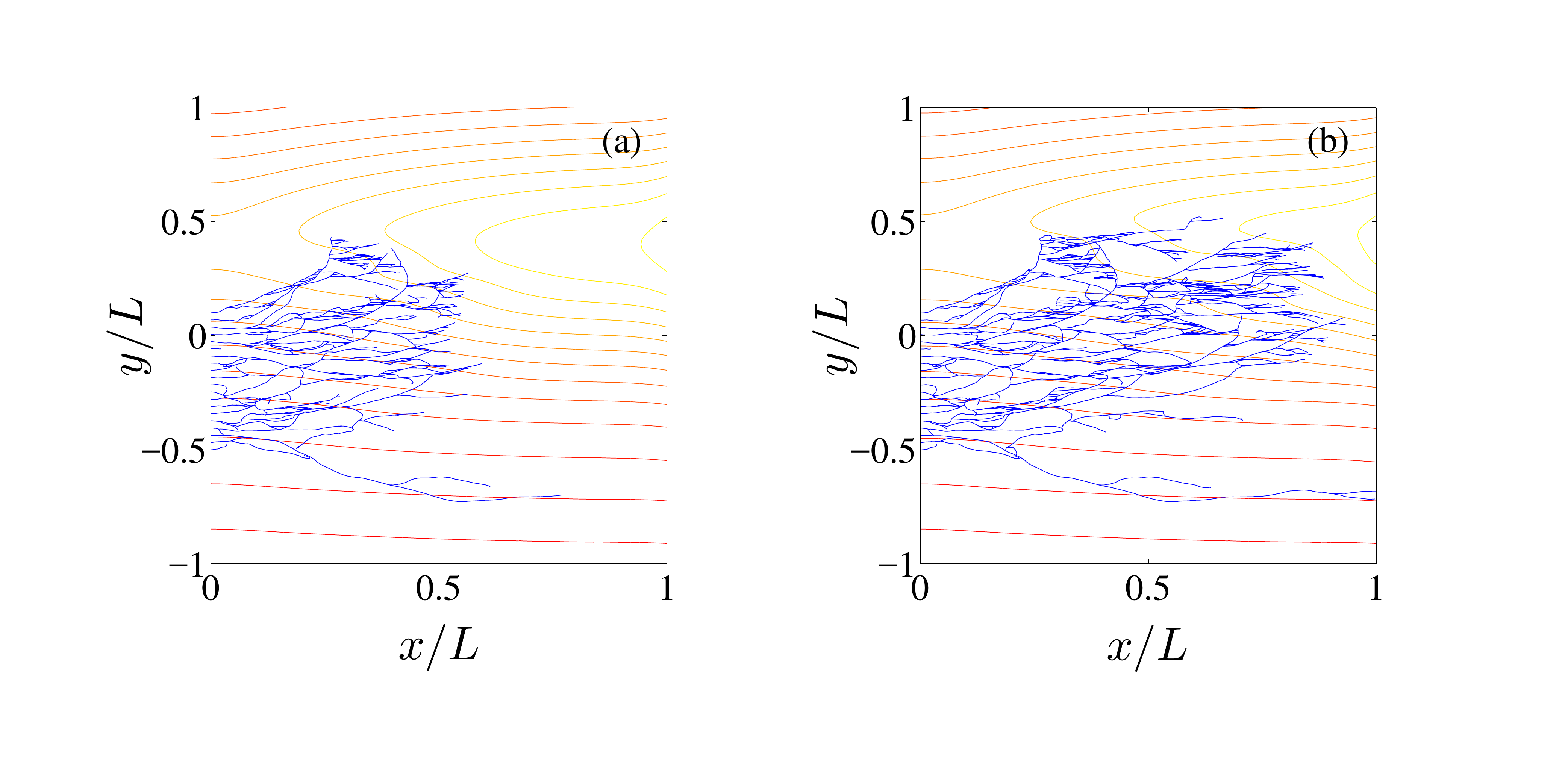}
\end{center}
\vskip-6mm 
\caption{Angiogenic network generated by the stochastic process at (a) 24 hr and (b) 36 hr after sprouts issue from the primary blood vessel at $x = 0$. The level curves of the TAF density $C(t,\mathbf{x})$ are also depicted, showing a tumor located vertically at $x=L$ above the $x$-axis. \label{fig1}}
\end{figure}

Previously, we have derived a deterministic description from a simple two dimensional (2D) tip cell model of tumor induced angiogenesis \cite{bon14}. This model considers tip cells subject to chemotactic, friction and white noise forces, and to random branching. When a moving sprout tip meets an existing sprout or a blood vessel, it fuses with it and stops moving, which is a simple model of anastomosis. A slightly more complicated earlier model by Capasso and Morale also includes haptotaxis \cite{cap09}. Fig.~\ref{fig1} shows two snapshots of a realization of the stochastic process. Our deterministic description \cite{bon14} consists of an integropartial differential equation for the density of active tip cells coupled to a reaction-diffusion equation for the tumor angiogenic factor (TAF), which is representative of VEGF and other relevant growth factors. Branching and anastomosis processes appear as source and sink terms in the equation for the density of active tip cells. The tip density and other mesoscopic quantities are ensemble averages over replicas of the stochastic process \cite{ter16}. A similar equation for the tip density on $\mathbb{R}^D$ ($D=2,3$) can be rigorously derived from the stochastic equations in the limit as the initial number of tips goes to infinity \cite{cap19}. However, when we consider the more realistic situation of angiogenesis issuing from a primary vessel and advancing in a bounded region, the number of tips is finite and limited by anastomosis. In this situation, the derivation of the deterministic equation from ensemble averages of the stochastic process makes more physical sense \cite{ter16}, although a rigorous proof of its validity seems more difficult.  

Analysis and numerical solutions of both deterministic and stochastic descriptions of angiogenesis show that the tip density advances from primary blood vessel to tumor as a moving lump, which is a {\em two dimensional diffusive soliton} (2DDS). The 2DDS is a solution of a simplified version of the integropartial equation for the density of active tips moving on $\mathbb{R}^2$ and with constant TAF concentration. In simple one dimensional (1D) geometries, the longitudinal profile of the moving lump is a one dimensional diffusive soliton (1DDS) \cite{bon16}. Diffusive solitons are stable solitary waves of dissipative systems which, unlike true solitons, do not emerge unchanged from collisions \cite{rem99}. How does the soliton picture apply to angiogenesis starting from a blood vessel and advancing toward a tumor? The distance between the primary vessel and the tumor has to be sufficiently large, for otherwise the 2DDS does not have space and time to form. Provided the distance is large enough, there are three stages for the motion of active tips. First, the active tips proliferate through branching until the 2DDS forms. Secondly, the 2DDS advances far from primary vessel and tumor. This second stage can be approximated by the 2DDS solution obtained for the case of infinite space and constant TAF concentration if the latter changes slowly. The velocity and shape of the soliton vary slowly to accommodate the changes in TAF concentration and are determined by collective coordinate equations (CCEs) \cite{bon16,bon16pre,bon17}. The last stage describes how the angiogenic network reaches the tumor. 

The main result of this work is finding and validating an approximate description and CCEs for the 2DDS in the general case. Note that angiogenesis is a biological process very far from equilibrium. However, numerical simulations show that when the density of active vessel tips is far from the boundaries (primary vessel and tumor), it evolves rapidly to a particular pattern, the 2DDS or {\em angiton}. The latter is a stable uniformly moving solution for a Fokker-Planck equation with source terms and constant TAF concentration in the spatially unbounded case. It plays the same role as the thermal equilibrium solution for the Fokker-Planck equation describing a system with detailed balance. The 2DDS is characterized by a few parameters (velocity, height). For a finite geometry and slowly-varying TAF concentration, the 2DDS parameters change slowly to accommodate the angiton motion towards the boundaries and the varying TAF concentration. 

The description of the second stage of angiogenesis is far from obvious. The first step is to reduce the 2D equation for the marginal tip density to a 1D equation. For equations deriving from a variational principle, such as the Gross-Pitaevskii equation for a cigar shaped Bose condensate, an appropriate Ansatz is a Gaussian function of the transversal coordinate times a function of the axial coordinate \cite{sal02}. Inserting this Ansatz into the action of the variational principle produces an equation for the longitudinal function and an equation for the width of the Gaussian. When the latter is solved in terms of the longitudinal function, it yields an effective 1D nonlinear equation for it \cite{sal02}. 

The deterministic equation governing tumor induced angiogenesis does not derive from a variational principle, and, therefore, we cannot use the same ideas to reduce it to a 1D equation. For a general configuration, the 2DDS does not move on a straight line and, therefore, we have to use curvilinear coordinates to characterize both the 2DDS and the trajectory of its center of mass. The longitudinal coordinate is directed along the instantaneous velocity of the 2DDS and the transversal coordinate is measured perpendicularly to the velocity. By using the method of multiple scales with a fast transversal length scale that characterizes the 2DDS width, (i) we show that the marginal tip density has Gaussian shape (with a small variance), and (ii) derive an averaged 1D equation for the 2DDS longitudinal profile. The latter equation is the same as derived earlier for a simple 1D geometry \cite{bon16pre} except for a renormalized anastomosis coefficient and motion over the longitudinal curvilinear coordinate. The solution of the 1D is a 1DDS. While the soliton is a stable traveling wave moving rigidly on the infinite the real line, the slow evolution of the TAF concentration and the influence of boundary conditions change the 2DDS shape and velocity. We derive equations for collective coordinates of the 2DDS that include the magnitude and orientation of the soliton velocity, its shape parameter, and the location of its center of mass. The 2DDS rapidly adjusts its shape to the instantaneous values of the collective coordinates. From the numerical solutions of the CCEs, we can reconstruct the  2DDS, which then yields the marginal density of active tips. Comparison with the marginal density obtained from ensemble averages shows that the 2DDS provides a good approximation for the stochastic description of the tip cell model. Possible future applications of the present work to biology include investigating control of the 2DDS motion, e.g. by studying the effect of antiangiogenic drugs on its dynamics; cf Ref.~\cite{lev01} for modifications of angiogenesis models due to angiostatin.

The rest of the paper is as follows. We recall the reduced integropartial differential equation for the marginal density of active vessel tips \cite{ter16} in Section \ref{sec:model}. To describe the lump of active tips moving toward the tumor, we assume that it is initially a Gaussian function of the transversal coordinate, check that it continues evolving as a Gaussian, derive an equation for its longitudinal part and find a one dimensional soliton as an approximate solution in Section \ref{sec:2DDS}. The analytical formula for the soliton is analogous to that found in \cite{bon16,bon16pre}. Section \ref{sec:cc} contains a derivation of the differential equations for the collective coordinates of the 2DDS. The coefficients of the CCEs are spatial averages over the TAF density. The width of narrow 2DDSs is time independent and only three collective coordinates are needed to describe them. In Section \ref{sec:numerical}, we explain how to calculate the coefficients in the CCEs, solve them numerically, reconstruct the 2DDS and, through it, the marginal vessel tip density. We compare it with ensemble averages of the stochastic process. It turns out that the location of the 2DDS peak approximately gives the location of the maximum of the marginal density for any realization of the stochastic process. Thus, the 2DDS roughly describes the advancing vessel network for each realization although different realizations provide different looking vessel networks. The conclusions of this work appear in Section \ref{sec:conclusions}. The Appendices are devoted to technical matters.  

\section{Angiogenesis model}\label{sec:model}
\subsection{Stochastic model}
Early stages of tumor induced angiogenesis are described by a simple stochastic model of motion, creation and destruction of active tips \cite{bon14,ter16}. The model comprises a system of Langevin-Ito equations for the extension of vessel tips, branching of tips (a birth process) and anastomosis (an annihilation process when the tips merge with existing vessels and cease to be active). In the model, the growing vessels are the trajectories of active vessel tips, whose motion obeys the following Langevin-Ito equations
\begin{eqnarray} 
d\mathbf{X}^i(t)&=&\mathbf{v}^i(t)\, dt,\nonumber\\
d\mathbf{v}^i(t)&=& \!\left[- \mathbf{v}^i(t)+\mathbf{F}\!\left(C(t,\mathbf{X}^i(t))\right)\right]\!\beta dt + 
\sqrt{\beta}\, d\mathbf{W}^i(t), \label{eq1}
\end{eqnarray}
where the $\mathbf{X}^i(t)$ and $\mathbf{v}^i(t)$ are the position and velocity of the $i$th tip at time $t$, respectively, the $\mathbf{W}^i(t)$ are independent identically distributed (i.i.d.) standard Brownian motions, and 
\begin{eqnarray}
\mathbf{F}(C)= \frac{\delta}{\beta}\,\frac{\nabla_x C(t,\mathbf{x})}{[1+\Gamma_1C(t,\mathbf{x})]^q}= \frac{\delta\nabla_x[1+\Gamma_1C(t,\mathbf{x})]^{1-q}}{(1-q)\Gamma_1\beta}. \quad \label{eq2}
\end{eqnarray}
The equation for the TAF density $C(t,\mathbf{x})$ is
\begin{eqnarray} 
\frac{\partial}{\partial t}C(t,\mathbf{x})\!&\!=\!& \! \kappa \Delta_x C(t,\mathbf{x})
-\chi C(t,\mathbf{x}) 
\sum_{i=1}^{N(t)} |\mathbf{v}^i(t)|\, \delta_{\sigma_x}(\mathbf{x}-\mathbf{X}^i(t)),  \label{eq3}
\end{eqnarray}
where $N(t)$ is the number of active tips at time $t$ and $\delta_{\sigma_x}$ are regularized delta functions:
\begin{eqnarray}
\delta_{\sigma_x}(\mathbf{x})= \frac{e^{-x^2/\sigma_x^2}\, e^{-y^2/\sigma_y^2}}{\pi\sigma_x \sigma_y}. \label{eq4}
\end{eqnarray}

The $i$th active tip branches and fuses (anastomoses) with another tip or vessel at random times $T^i$ and $\Theta^i$, respectively. At time $T^i$ a new active tip is created, whereas at time $\Theta^i$ the $i$th tip merges with a vessel or reaches the tumor, thereby ceasing to be active and counting as such. We ignore branching from mature vessels or from existing vessel sprouts; see \cite{cap19}. 

The probability that a tip branches from one of the existing ones during an infinitesimal time interval $(t, t + dt]$ is proportional to $\sum_{i=1}^{N(t)}\alpha(C(t,\mathbf{X}^i(t)))dt$, where $\alpha(C)$ is given by 
\begin{equation}
\alpha(C)=\frac{A\, C}{1+C},\quad A>0. \label{eq5}
\end{equation}
At time $T^i$, the velocity of the new tip that branches from tip $i$ is selected out of a normal distribution, $\delta_{\sigma_v}(\mathbf{v}-\mathbf{v}_0)$, with mean $\mathbf{v}_0$ and a narrow variance $\sigma_v^2$. The regularized delta function $\delta_{\sigma_v}(\mathbf{x})$ is given by Eq.~\eqref{eq4} with $\sigma_x=\sigma_y=\sigma_v$. 

The change per unit time of the number of tips in boxes $d\mathbf{x}$ and $d\mathbf{v}$ about $\mathbf{x}$ and $\mathbf{v}$ is
\begin{eqnarray}\nonumber
&&\sum_{i=1}^{N(t)}\alpha(C(t,\mathbf{X}^i(t)))\, \delta_{\sigma_v}(\mathbf{v}^i(t)-\mathbf{v}_0)=\int_{d\mathbf{x}}\int_{d\mathbf{v}}\alpha(C(t,\mathbf{x)})\\
&&\times \delta_{\sigma_v}(\mathbf{v}-\mathbf{v}_0)\sum_{i=1}^{N(t)}\delta(\mathbf{x}-\mathbf{X}^i(t))\delta(\mathbf{v}-\mathbf{v}^i(t)) d\mathbf{x} d\mathbf{v}. \label{eq6}
\end{eqnarray}
Representative values of all involved dimensionless parameters can be found in Table \ref{table2} \cite{bon14,ter16}.
\begin{table}[ht]
\begin{center}\begin{tabular}{cccccccccc}
 \hline
$\delta$ & $\beta$ &$A$& $\Gamma$& $\Gamma_1$ &$\kappa$&$\chi$&$\sigma_v$& $\sigma_x$ & $\sigma_y$\\
 $\frac{d_1C_R}{\tilde{v}_0^2}$ & $\frac{kL}{\tilde{v}_0}$ & $\frac{\alpha_1L}{\tilde{v}_0^3}$ & $\frac{\gamma}{\tilde{v}_0^2}$&$\gamma_1C_R$& $\frac{d_2}{\tilde{v}_0 L}$& $\frac{\eta}{L}$&-&- &-\\
1.5 & 5.88 & $22.42$ & 0.189 & 1& $0.0045$ & 0.002 & 0.08 &0.15 &0.05\\
 \hline
\end{tabular}
\end{center}
\caption{Dimensionless parameters and representative values. } 
\label{table2}
\end{table}

\subsection{Deterministic equations}
It is possible to derive deterministic equations for the density of {\em active} vessel tips and the vessel tip flux from the stochastic model. In all cases, the law of large numbers \cite{gar10} is involved. As the initial number of active tips $N(0)$ tends to infinity, the {\em scaled tip density}, defined as the number of active tips per unit phase volume divided by $N(0)$, is a self-averaging quantity obeying a deterministic integrodifferential equation \cite{cap19}, similar to that derived earlier in \cite{bon14}. This can be proved rigorously as an initial value problem for tips moving on the infinite space \cite{cap19}. However, the situation is different for a slab geometry in two space dimensions, where tips are born from a primary vessel and advance toward a tumor placed at a finite distance. In this case, proliferation of active vessels due to branching is balanced by their inactivation due to anastomosis, which typically keeps the number of active tips below one hundred \cite{ter16}. Numerical simulations of the stochastic process also indicate that there are substantial velocity and density fluctuations. Thus, there is numerical evidence that the density of active tips is not self-averaging for bounded geometries. 

An average density satisfying a deterministic equation can be determined by a different usage of the law of large numbers. We consider a large number $\mathcal{N}$ of replicas (realizations) $\omega$ of the stochastic process and introduce the following empirical averages \cite{ter16}:
\begin{eqnarray}
p_{\mathcal{N}}\!(t,\mathbf{x},\mathbf{v})\!&=&\!\frac{1}{\mathcal{N}}\sum_{\omega=1}^\mathcal{N}\sum_{i=1}^{N(t,\omega)}\delta_{\sigma_x}(\mathbf{x}-\mathbf{X}^i(t,\omega))
\delta_{\sigma_v}(\mathbf{v}-\mathbf{v}^i(t,\omega)),\label{eq7}\\
\tilde{p}_{\mathcal N}(t,\mathbf{x})\!\!&=&\!\frac{1}{\mathcal{N}}\sum_{\omega=1}^\mathcal{N}\sum_{i=1}^{N(t,\omega)}\delta_{\sigma_x}(\mathbf{x}-\mathbf{X}^i(t,\omega)), \label{eq8}\\
j_{\mathcal N}(t,\mathbf{x})\!\!&=&\!\frac{1}{\mathcal{N}}\!\sum_{\omega=1}^\mathcal{N}\!\sum_{i=1}^{N(t,\omega)}\!\!|\mathbf{v}^i(t,\omega)|
\delta_{\sigma_x}(\mathbf{x}-\mathbf{X}^i(t,\omega)).\label{eq9}
\end{eqnarray}
Replicas of the stochastic process are independent by definition. By the law of large numbers as $\mathcal{N}\to\infty$, the limits of these empirical averages become the expected values of the tip density, the marginal tip density and the vessel tip flux according to the stochastic process \cite{gar10}. The averages over the set of replicas are the usual ensemble averages of statistical mechanics. The densities in Eqs.~\eqref{eq8} and \eqref{eq9} are not probability densities: note that the integral of the marginal density $\tilde{p}_{\mathcal N}(t,\mathbf{x})$ over space is not one. Instead, it is the average number of active tips at time $t$, $\langle\langle N(t)\rangle\rangle=\mathcal{N}^{-1}\sum_{\omega=1}^\mathcal{N}N(t,\omega)$, which is finite in the limit of infinitely many replicas and changes with time due to the branching and anastomosis processes. 

In Ref.~\cite{ter16}, we have shown that the ensemble averages, $p(t,\mathbf{x},\mathbf{v})$, $\tilde{p}(t,\mathbf{x})$, and $j(t,\mathbf{x})$, solve the following equations: 
\begin{eqnarray}
&&\frac{\partial}{\partial t} p(t,\mathbf{x},\mathbf{v})=\alpha(C(t,\mathbf{x}))\,
 p(t,\mathbf{x},\mathbf{v})\delta_{\sigma_v}(\mathbf{v}-\mathbf{v}_0)
- \Gamma\, p(t,\mathbf{x},\mathbf{v}) \!\int_0^t\! \tilde{p}(s,\mathbf{x})\, ds \nonumber\\
&&\!- \mathbf{v}\!\cdot\! \nabla_x   p(t,\mathbf{x},\mathbf{v})  -\beta \nabla_v\! \cdot [(\mathbf{F}(C(t,\mathbf{x}))-\mathbf{v}) p(t,\mathbf{x},\mathbf{v})] 
+ \frac{\beta}{2} \Delta_{v} p(t,\mathbf{x},\mathbf{v}),
\label{eq10}\\
&& \tilde{p}(t,\mathbf{x})=\int p(t,\mathbf{x},\mathbf{v}')\, d \mathbf{v'}. \label{eq11}
\end{eqnarray}
Here, the first two terms on the right-hand side of Eq.~\eqref{eq10} describe branching and anastomosis, respectively. The other terms are the usual ones appearing in the Fokker-Planck equation corresponding to the Langevin equations \eqref{eq1}. The TAF equation \eqref{eq3} becomes
\begin{eqnarray} 
\frac{\partial}{\partial t}C(t,\mathbf{x})=\kappa \Delta_x C(t,\mathbf{x})- \chi\, C(t,\mathbf{x})\, j(t,\mathbf{x}),\label{eq12}
\end{eqnarray}
where 
\begin{equation}
j(t,\mathbf{x})= \int |\mathbf{v}'|\, p(t,\mathbf{x},\mathbf{v}')\, d \mathbf{v'}. \label{eq13}
\end{equation}
Unlike the case of the mean field limit $N(0)\to\infty$ in Ref.~\cite{cap19}, we lack a proof that the considered ensemble averages obey the deterministic equations \eqref{eq10}-\eqref{eq12}. However, we can compare the solution of the deterministic equations to the ensemble averages of the stochastic process and fit the anastomosis coefficient $\Gamma$ so that the average number of tips as a function of time is as close as possible in both descriptions \cite{ter16,bon16pre}. The number of replicas $\mathcal{N}$ in our numerical simulations is selected so that the ensemble averages, $p_{\mathcal N}(t,\mathbf{x},\mathbf{v})$, $\tilde{p}_{\mathcal N}(t,\mathbf{x})$, and $j_{\mathcal N}(t,\mathbf{x})$, do not change by adding any more replicas to $\mathcal{N}$. Appendix \ref{app1} contains appropriate initial and boundary conditions for solving the deterministic equations \eqref{eq10} and \eqref{eq12}. 

In principle, the ensemble average view of angiogenesis could be used to calculate higher moments, not only averaged quantities. This is largely unexplored. In Ref.~\cite{ter16}, we have used Ito's formula with added branching and anastomosis to obtain an equation for the tip density. To this end, we need some closure assumptions of the type $\langle\langle f(x)\rangle\rangle= f(\langle\langle x\rangle\rangle)$. Justifying these assumptions would require taking into account and analyzing density fluctuations. This could be done by deriving a hierarchy of equations for $n$-particle densities and using closure assumptions as in kinetic theory \cite{bog46,cer69,akh81}. We could also include density fluctuations by keeping a Poisson noise (representing random branching) in Eq.~\eqref{eq10} for the active tip density \cite{ter16}. This would give a formulation akin to the fluctuating lattice Boltzmann equation \cite{dun07}. In other cases, such as  in the classical statistical mechanics of a crystal \cite{lew61} or in fluid turbulence \cite{lew62}, it has been possible to derive and analyze functional equations. In recent years, there has been much progress in understanding rigorously the Kolmogorov-Hopf functional differential equation for fluid turbulence and the underlying invariant measure, \cite{bir13}. 

\subsection{Marginal tip density} 
Assuming that the extension of the moving angiogenic sprouts is overdamped, Eqs.~\eqref{eq10} and \eqref{eq12} yield the following system of nondimensional equations for the marginal density of active vessel tips, $\tilde{p}(t,\mathbf{x})$, and the TAF density, $C(t,\mathbf{x})$, \cite{bon16pre}
\begin{eqnarray}
&&\frac{\partial\tilde{p}}{\partial t}+\nabla_x\!\cdot\!(\mathbf{F}\tilde{p})-\frac{1}{2\beta}\Delta_x\tilde{p}=\mu\,\tilde{p}
-\Gamma\tilde{p}\!\int_0^t\!\tilde{p}(s,\mathbf{x}) ds, \quad\label{eq14}\\
&&\frac{\partial}{\partial t}C(t,\mathbf{x})=\kappa \Delta_x C(t,\mathbf{x})- \tilde{\chi}\, C(t,\mathbf{x})\,\tilde{p}(t,\mathbf{x}).\label{eq15}
\end{eqnarray}
Here
\begin{eqnarray}
&&\mu=\frac{\alpha}{\pi}\left[1+\frac{\alpha}{2\pi\beta(1+\sigma_v^2)}\ln\!\left(1+\frac{1}{\sigma_v^2}\right)\!\right]\!,\quad \tilde{\chi}=J\chi,\quad\nonumber\\
&&J=\int_0^\infty\! dV\frac{Ve^{-V^2}}{\pi}\!\int_{-\pi}^\pi\!d\phi\sqrt{1+V^2+2V\cos\phi}. \label{eq16}
\end{eqnarray}
Note that the details of velocity selection during branching are lumped in the function $\mu(C)$ given by Eq.~\eqref{eq16}. In Ref.~\cite{bon16pre}, Eqs.~\eqref{eq14} and \eqref{eq15} are derived from Eqs.~\eqref{eq10} and \eqref{eq12} by a Chapman-Enskog perturbation method that shows the tip density to be $p\sim e^{-|\mathbf{V}|^2}\tilde{p}(t,\mathbf{x})/\pi$. Inserting this density in Eq.~\eqref{eq13} and replacing $\mathbf{v}'=\mathbf{v}_0+\mathbf{V}$, we find $j(t,\mathbf{x})= J\tilde{p}(t,\mathbf{x})$ ($J\approx 1.28192$).

\begin{table}[ht]
\begin{center}\begin{tabular}{cccccc}
 \hline
$\mathbf{x}$& $\mathbf{v}$ & $t$ &$C$& $p$ &$\tilde{p}$\\ 
$L$ & $\tilde{v}_0$ & $\frac{L}{\tilde{v}_0}$ & $C_R$ & $\frac{1}{\tilde{v}_0^2L^2}$& $\frac{1}{L^2}$\\ 
mm&$\mu$m/hr & hr& mol/m$^2$&$10^{21}$s$^2$/m$^4$ & $10^{5}$m$^{-2}$\\
$2$& 40 & 50 & $10^{-16}$ & 2.025 & 2.5 \\
\hline
\end{tabular}
\end{center}
\caption{Dimensional units with representative values. }
\label{table1}
\end{table}

Typical values of the positive dimensionless parameters appearing in Eqs.~\eqref{eq14}-\eqref{eq15} are given in Table \ref{table2} whereas Table \ref{table1} gives the units used to nondimensionalize them \cite{bon14}. Table \ref{table2} shows that $1/\beta$, $\kappa$ and $\chi$ are small. This means that the evolution of the TAF density is slow compared to that of $\tilde{p}$ and that the diffusion in Eq.~\eqref{eq14} is small.

\begin{figure}[h]
\begin{center}
\includegraphics[width=14cm]{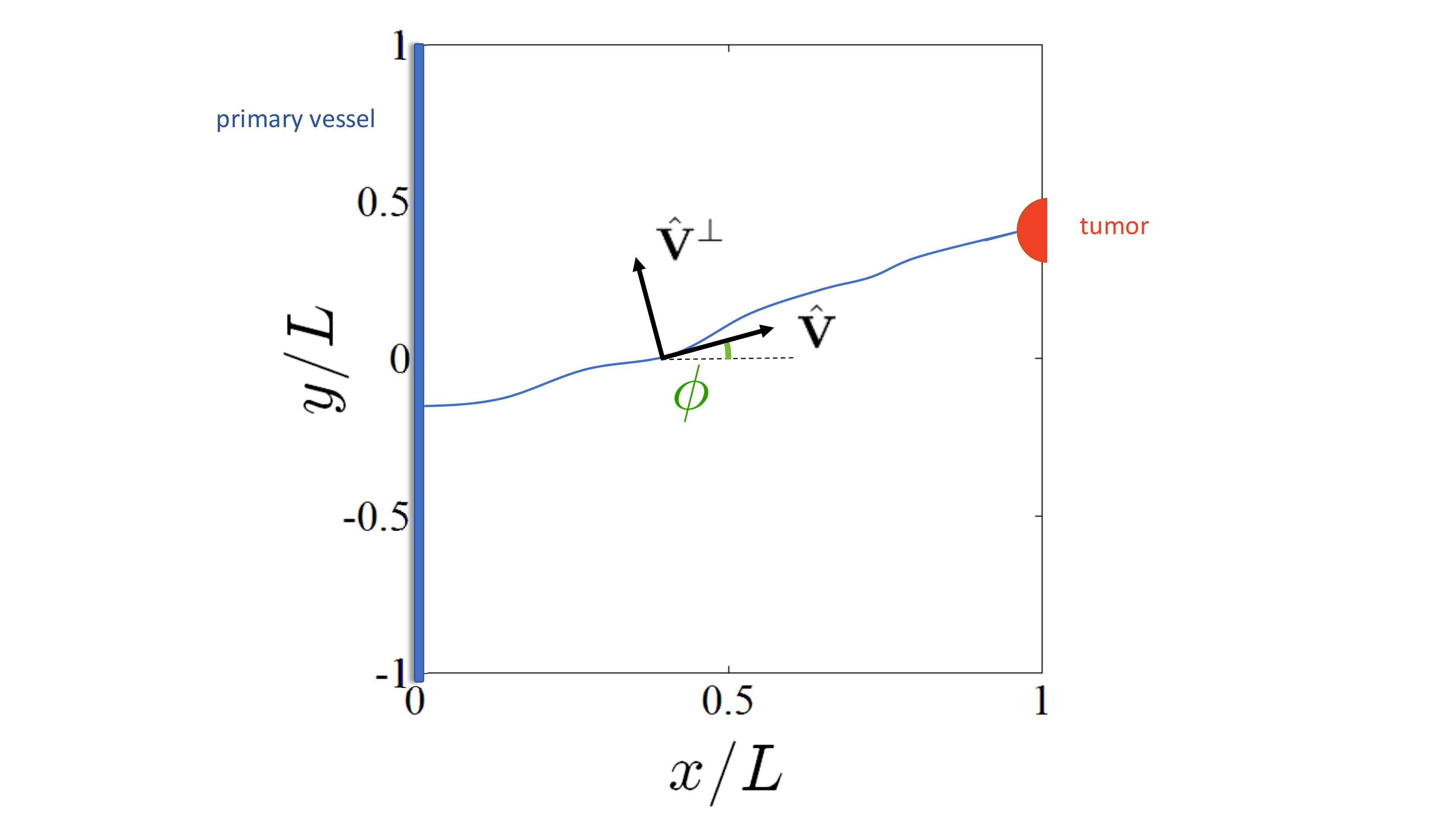} 
\end{center}
\vskip-3mm 
\caption{ Sketch of the unit vectors $\hat{\mathbf{V}}=(\cos\phi,\sin\phi)$ and $\hat{\mathbf{V}}^\perp=(-\sin\phi,\cos\phi)$. \label{fig2}}
\end{figure}

\section{Two dimensional diffusive soliton}
\label{sec:2DDS}
In this section, we derive the approximate 2DDS by using a method of multiple scales.

\subsection{Marginal tip density in curvilinear coordinates}
We now find an approximate lump-shaped solution of Eq.~(\ref{eq14}). Let $\mathbf{X}(t)$ be the center of mass of the lump at time $t$. Longitudinal and transverse coordinates based on the trajectory $\mathbf{X}(t)$ are
\begin{eqnarray}
&&\xi=(\mathbf{x}-\mathbf{X})\!\cdot\!\hat{\mathbf{V}}, \quad\eta=(\mathbf{x}-\mathbf{X})\!\cdot\!\hat{\mathbf{V}}^\perp,\label{eq17}\\
&&\mathbf{V}=\dot{\mathbf{X}}=c\hat{\mathbf{V}}=c(\cos\phi,\sin\phi), \quad
\hat{\mathbf{V}}^\perp=(-\sin\phi,\cos\phi).\label{eq18}
\end{eqnarray}
See Fig.~\ref{fig2}. Here and henceforth, $\dot{f}(t)=df/dt$ for any function of time, $c=|\mathbf{V}|$. Thus, $\mathbf{x}=(x,y)=(X,Y) + (\dot{X},\dot{Y})\,\xi/c+(-\dot{Y},\dot{X})\,\eta/c$. Eq.~\eqref{eq14} can be written as
\begin{eqnarray}
&&\frac{\partial\tilde{p}}{\partial t}+\frac{\partial}{\partial\xi}\!\left((F_\xi-c+\eta\dot{\phi})\tilde{p}-\frac{1}{2\beta}\frac{\partial\tilde{p}}{\partial\xi}\!\right)\! 
+\frac{\partial}{\partial\eta}\!\left((F_\eta-\xi\dot{\phi})\tilde{p}-\frac{1}{2\beta}\frac{\partial\tilde{p}}{\partial\eta}\!\right)\!-\mu\,\tilde{p} \nonumber\\
&&=-\Gamma\tilde{p}\!\int_0^t\!\tilde{p}(s,\xi(s),\eta(s))\, ds, \label{eq19}\\
&&F_\xi=\hat{\mathbf{V}}\cdot\mathbf{F},\quad F_\eta=\hat{\mathbf{V}}^\perp\cdot\mathbf{F},\label{eq20}
\end{eqnarray}
because $\dot{\xi}=-c+\eta\dot{\phi}$ and $\dot{\eta}=-\xi\dot{\phi}$.

\subsection{Method of multiple scales}
We shall now assume that the initial TAF concentration is a Gaussian with a small variance across the transversal direction $\eta$. We assume that $\tilde{p}$ depends on a fast variable $\eta/\sigma$ and a slowly varying $\eta$. We will use a method of multiple scales to find the slowly varying part of the marginal tip density \cite{kev96,BT10}. The dominant terms in Eq.~\eqref{eq19} are
\begin{subequations}\label{eq21}
\begin{eqnarray}
\frac{\partial}{\partial\eta}(F_\eta\tilde{p})\sim\frac{1}{2\beta}\frac{\partial^2\tilde{p}}{\partial\eta^2}\Longrightarrow\tilde{p}\sim e^{2\beta\Omega[C]}\Psi(t,\xi),\label{eq21a}\\
\Omega[C]=\frac{\delta}{\beta}\frac{(1+\Gamma_1C)^{1-q}}{\Gamma_1(1-q)}\sim \Omega^0(\eta) - \frac{\eta^2}{2}\left.\!\left|\!\left(\frac{\partial^2 \Omega}{\partial\eta^2}\right)\!\right|_{\eta=0}\right|\!, \label{eq21b}
\end{eqnarray}
\end{subequations}
in which $\partial\Omega/\partial\eta= F_\eta=0$ at $\eta=0$, where $C$ reaches its local maximum. Then the approximate density is a narrow Gaussian in $\eta$ that satisfies 
\begin{subequations}\label{eq22}
\begin{eqnarray}
&&\tilde{p}(t,\mathbf{x})\sim\frac{e^{-\eta^2/\sigma^2}}{\sqrt{\pi}\,\sigma} P(t,\xi,\eta)\sim\delta(\eta)P(t,\xi,0),\quad \label{eq22a}\\&&
\sigma^2=\left.\frac{1}{\beta|\frac{\partial^2 \Omega}{\partial\eta^2}|}\right|_{\eta=0}=\left.\frac{[1+\Gamma_1C]^q}{\left|\frac{\partial^2C}{\partial\eta^2}\right|\delta}\right|_{\eta=0},\label{eq22b}
\end{eqnarray}\end{subequations}
as $\sigma\to 0$. We have identified $\sigma$ and, for the numerical values in Table \ref{table2}, $\sigma$ is indeed small; see Section \ref{sec:numerical}. Then $P(t,\xi,\eta)$ varies slowly in $\eta$ compared to the Gaussian prefactor in Eq.~\eqref{eq22}. To obtain a reduced equation for $P$, we multiply Eq.~\eqref{eq19} by $\sigma$ and integrate it with respect to the fast variable $\tilde{\eta}=\eta/\sigma$. After this, we use the second approximation in Eq.~\eqref{eq22a} and set the slow variable $\eta=0$. The resulting equation for $P(t,\xi,0)$ is  
\begin{eqnarray}
&&\frac{\partial P}{\partial t}+\frac{\partial}{\partial\xi}\!\left((\overline{F_\xi}-c)P-\left.\frac{1}{2\beta}\frac{\partial P}{\partial\xi}\!\right)\right|_{\eta=0}\!+\frac{\partial}{\partial\eta}\!\left((\overline{F_\eta}-\xi\dot{\phi})P-\left.\frac{1}{2\beta}\frac{\partial P}{\partial\eta}\!\right)\right|_{\eta=0}\nonumber\\
&&\quad=\overline{\mu}\, P -\frac{\Gamma}{\sigma\sqrt{2\pi}}\, P \int_0^t\!P(s,\xi,0)\, ds,\label{eq23}
\end{eqnarray}
\begin{subequations}\label{eq24}
\begin{eqnarray}
&&\overline{f(\tilde{\eta},\eta)}= \frac{1}{\sqrt{\pi}}\int_{-\infty}^\infty f(\tilde{\eta},0)\, e^{-\tilde{\eta}^2} d\tilde{\eta},\label{eq24a}\\
&& \sigma\int_{-\infty}^\infty\frac{\partial}{\partial\eta}[f(\tilde{\eta},\eta)\tilde{p}]\,d\tilde{\eta} =\left. \frac{\partial}{\partial\eta}[P\,\overline{f(\tilde{\eta},\eta)}]\right|_{\eta=0},\label{eq24b}\\
&&\sigma\int_{-\infty}^\infty\tilde{p}\, d\tilde{\eta}=P(t,\xi,0). \quad\label{eq24c}
\end{eqnarray}
\end{subequations}

\subsection{One dimensional soliton}
Keeping Eq.~\eqref{eq22a} in mind, we will now look for a solitary wave profile of $P(t,\xi,0)$ as in Refs.~\cite{bon16,bon16pre}, which, by abuse of language, we shall call {\em soliton}. The coefficients $\kappa$ and $\chi$ in Eq.~(\ref{eq15}) are very small \cite{bon14} and therefore the TAF concentration varies very slowly compared with the marginal tip density. When writing  Eq.~\eqref{eq23}, we have used that $F_\xi$ and $F_\eta$ depend on $C$ and, therefore, vary slowly in time and space. Note that the slow variation of $C$ implies that an initial Gaussian TAF concentration with small variance produces an initial $\sigma\ll 1$, which does not change in time. Thus, Eq.~\eqref{eq22} may hold initially due to a peaked initial $C$ and persist in time even if the coefficients in Eq.~\eqref{eq19} are of order 1.

For time and $\eta$ independent $P=P(\xi)$, with $dt=-d\xi/c$, we now define
\begin{equation}
\rho(\xi)=-\frac{1}{c}\int_{\xi(0)}^\xi P(\xi')\, d\xi', \label{eq25}
\end{equation}
insert it in Eq.~\eqref{eq23} and integrate with respect to $\xi$. The result is 
\begin{eqnarray}
\frac{\partial}{\partial\xi}\!\left(\!(\overline{F_\xi} - c)\rho-\frac{1}{2\beta}\frac{\partial\rho}{\partial\xi}\!\right)\! 
\!=\!\overline{\mu} \rho-\frac{g\rho^2}{2}+K,\quad g\!=\!\frac{\Gamma}{\sigma\sqrt{2\pi}},\, \label{eq26}
\end{eqnarray}
in which $K$ is independent of $\xi$ and $g$ is a {\em renormalized} anastomosis coefficient. We will also assume that the initial TAF concentration varies on a larger spatial scale than the soliton size, which constitutes a good approximation \cite{bon14}. Then $\overline{F_\xi}$ and $\overline{\mu}$ are almost constant. Ignoring diffusion, we obtain
\begin{eqnarray}
\left(c-\overline{F_\xi}\right)\!\frac{2}{g}\frac{\partial\rho}{\partial\xi}=\rho^2-2\frac{\overline{\mu}}{g}\rho-\frac{2K}{g}. 
\label{eq27}
\end{eqnarray}
Setting $\rho=\frac{\overline{\mu}}{g}+\nu\tanh(\lambda\xi)$, we find $\nu^2=\frac{\overline{\mu}^2+2Kg}{g^2}$ and $2\nu\lambda(c-\overline{F_\xi})/g=-\nu^2$, thereby obtaining
\begin{eqnarray}
\rho=\frac{\overline{\mu}}{g}-\frac{\sqrt{2Kg+\overline{\mu}^2}}{g}\tanh\!\left[\frac{\sqrt{2Kg+\overline{\mu}^2}}{2(c-\overline{F_\xi})}\xi\right]\!.  \label{eq28}
\end{eqnarray}
Here a constant of integration has been absorbed in the definition of $\xi$. Thus $P=-c\partial\rho/\partial\xi$ yields \cite{bon16,bon16pre}
\begin{eqnarray}
P_s(\xi)=\frac{(2Kg+\overline{\mu}^2)c}{2g(c-\overline{F_\xi})}\mbox{sech}^2\!\left[\frac{\sqrt{2Kg+\overline{\mu}^2}}{2(c-\overline{F_\xi})}\xi\right]\!.  \label{eq29}
\end{eqnarray}
As indicated in Refs.~\cite{bon16,bon16pre}, Eq.~\eqref{eq29} is similar to the usual soliton solution of the Korteweg-de Vries equation, except that the soliton velocity and shape now depend on three parameters, $c$, $K$, and (implicitly through $\xi$) $\phi$. Note that the existence of the 2DDS solution of Eqs.~\eqref{eq22} and \eqref{eq29} is a consequence of the quadratic anastomosis term in Eq.~\eqref{eq14} first derived in Ref.~\cite{bon14}. The function $\rho$ in Eq.~\eqref{eq25} is the integral of the square hyperbolic secant given by  Eq.~\eqref{eq29}, which has the same shape as the Korteweg-de Vries soliton \cite{lax68}. While the latter results from a balance of time derivative, nonlinear convection and dispersion \cite{lax68}, the 1DDS soliton of Eq.~\eqref{eq29} comes from a balance of time derivative, linear convection, branching and anastomosis (which contains a memory term).

\subsection{Center of mass}
Using Eqs.~\eqref{eq22a} and \eqref{eq29}, we can calculate the center of mass of the 2DDS in curvilinear coordinates. As $\tilde{p}(t,\mathbf{x})=e^{-\eta^2/\sigma^2}P_s(\xi)/(\sqrt{\pi}\sigma)$ is even in both $\eta$ and $\xi$, the center of mass is the origin: $\int (\xi,\eta)\,\tilde{p}(t,\mathbf{x})d\xi\, d\eta/\int \tilde{p}(t,\mathbf{x}) d\xi\, d\eta=(0,0)$. Thus, the center of mass of the 2DDS coincides with the peak of the marginal tip density when the soliton is a good approximation for the latter. 

\section{Collective coordinates}\label{sec:cc}
Numerical simulations suggest that the 2DDS solution moving on unbounded space is stable. To obtain the soliton formula \eqref{eq29}, we ignored the effects of small diffusion and a slowly varying TAF concentration. Without these terms, a proof that the traveling wavefront solution Eq.~\eqref{eq28} is linearly stable (up to uniform spatial translations) follows along the same lines of Ref.~\cite{bon91}. We expect the effects of diffusion and the slow TAF evolution to make the collective coordinates $c$, $K$ and $\phi$ slowly varying functions of time: due to its stability, the 2DDS adjusts its shape and velocity to the instantaneous values of the collective coordinates. 

For equations deriving from a variational principle, such as the Gross-Pitaevskii equation for a cigar shaped Bose condensate, a derivation of the CCEs first assumes that the soliton is a Gaussian function of the transversal coordinate times a function of the axial coordinate \cite{sal02}. Then this Ansatz is inserted into the variational principle and the corresponding Euler-Lagrange equations are the CCEs. In our case, we do not have a variational principle. Instead, the method of multiple scales has provided us with the splitting of the 2DDS in a Gaussian of the transversal coordinate times the longitudinal 1D diffusive soliton, cf Eq.~\eqref{eq22a}. What is an Ansatz in Ref.~\cite{sal02} is provided by our theory of the 2DDS. 

\subsection{Slow variations of the collective coordinates}
To obtain the 2DDS evolution without recourse to a variational principle (which does not exist for the present problem), we observe the following. As $F_\xi$ and $\mu$ are functions of $C(t,\mathbf{x})$, $P_s$ is primarily a function of $\xi$, but it is also a slowly varying function of $\xi$, $\eta$ and $t$ through the $\tilde{\eta}$-averages of the coefficients $\overline{\mu}$ and $\overline{F_\xi}$. Then
\begin{equation}
P_s=P_s\!\left(\xi;K,c,\overline{\mu(C)},\overline{F_\xi\!\left(C,\frac{\partial C}{\partial\xi}\right)}\!\right)\!. \label{eq30}
\end{equation}
The averages over the fast transversal coordinate $\tilde{\eta}$ still vary rapidly with the longitudinal coordinate $\xi$ of Eq.~\eqref{eq29} and vary slowly on $\xi$ and $\eta$ through the  the TAF concentration, which varies smoothly with distance. As indicated in Appendix \ref{app3}, we shall consider that $\mu(C)$ is approximately constant and set $\partial C/\partial t=0$ because the TAF concentration is varying slowly with time. Then we have 
\begin{eqnarray}
&&P_s\!\!\left(\xi;K,c,\overline{\mu(C)},\overline{F_\xi\!\left(C,\frac{\partial C}{\partial\xi}\right)}\!\right)\!=P_s(\xi;K,c,\langle\overline{F_\xi}\rangle)\nonumber\\
&&\quad
+\,\frac{\partial P_s}{\partial\overline{F_\xi}}(\xi;K,c,\langle\overline{F_\xi}\rangle)\,(\overline{F_\xi}-\langle\overline{F_\xi}\rangle)\!+\!\ldots,    \label{eq31}
\end{eqnarray}
in which we have dropped the dependence of $P_s$ on $\mu(C)$ and expanded the averages over the fast transversal coordinate, $\overline{f(C(t,\xi,\eta))}$, to first order in their differences with spatial averages
\begin{eqnarray}
\langle f(C(t,\psi,\eta))\rangle =\frac{1}{b-a}\int_a^bf(\psi)\, d\psi. \label{eq32}
\end{eqnarray}
Here $f(\psi)\equiv \overline{f(C(t_0,\psi,0))}$. Eq.~\eqref{eq23} is an average over the longitudinal coordinate, in which we have ignored time variation of the TAF concentration after some time $t=t_0$ and set $\eta=0$, cf Eq.~\eqref{eq22a}. The time $t_0$ is selected after the formation stage of the 2DDS, and the interval $\mathcal{I}=(a,b)$ has to be appropriately chosen, as discussed in the next section. Using Eq.~\eqref{eq31}, we find
\begin{eqnarray}
\nabla_\xi P_s\!&=&\!(1,0)\,\frac{\partial P_s}{\partial\xi}(\xi;K,c,\langle\overline{F_\xi}\rangle)+ 
\frac{\partial P_s}{\partial\overline{F_\xi}}(\xi;K,c,\langle\overline{F_\xi}\rangle)\nabla_\xi\overline{F_\xi}+\ldots,\label{eq33}\\
\Delta_\xi P_s\!&=&\!\frac{\partial^2P_s}{\partial\xi^2}(\xi;K,c,\langle\overline{F_\xi}\rangle)+ 
\frac{\partial P_s}{\partial\overline{F_\xi}}(\xi;K,c,\langle\overline{F_\xi}\rangle)\Delta_\xi\overline{F_\xi} + \nonumber\\
&+&
2\frac{\partial^2P_s}{\partial\xi\partial\overline{F_\xi}}(\xi;K,c,\langle \overline{F_\xi}\rangle)\frac{\partial\overline{F_\xi}}{\partial\xi}+\ldots,\quad\label{eq34}
\end{eqnarray}
where $\nabla_\xi=(\partial/\partial\xi,\partial/\partial\eta)$ and $\Delta_\xi=\nabla^2_\xi$. We now insert Eq.~(\ref{eq30}) into Eq.~(\ref{eq23}) and use Eqs.~\eqref{eq31}, \eqref{eq33} and \eqref{eq34} to simplify the result, thereby obtaining (see Appendix \ref{app3})
\begin{eqnarray}
&&\frac{\partial P_s}{\partial K} \dot{K}\!+\!\frac{\partial P_s}{\partial c}\dot{c}+\!\left( \frac{\partial P_s}{\partial\langle\overline{F_\xi}\rangle} \overline{F_\eta} - \xi\frac{\partial\overline{F_\xi}}{\partial\eta}\frac{\partial P_s}{\partial\langle\overline{F_\xi}\rangle} \right)\!\dot{\phi}=\mathcal{A},\qquad\, \label{eq35}\\
&&\mathcal{A}=  \frac{1}{2\beta}\frac{\partial^2P_s}{\partial\xi^2} -\frac{\partial P_s}{\partial\langle\overline{F_\xi}\rangle}\!\left[\!\left((\overline{F_\xi}-c)\frac{\partial}{\partial\xi}+\overline{F_\eta}\frac{\partial}{\partial\eta}\right)\!\overline{F_\xi} 
-\frac{\Delta_\xi\overline{F_\xi}}{2\beta}\right]\! \nonumber\\ 
&&\quad-P_s\!\left(\frac{\partial\overline{F_\xi}}{\partial\xi}+\frac{\partial\overline{F_\eta}}{\partial\eta}\right)\!+\frac{1}{\beta}\frac{\partial\overline{F_\xi}}{\partial \xi}\frac{\partial^2P_s}{\partial\xi\partial\langle\overline{F_\xi}\rangle} + \frac{1}{2\beta}\frac{\partial^2 P_s}{\partial\langle\overline{F_\xi}\rangle^2} \,|\nabla_{\xi}\overline{F_\xi}|^2.  \quad\quad\label{eq36}
\end{eqnarray}
In these equations, $P_s=P_s(\xi;K,c,\langle\overline{F_\xi}\rangle)$. We now find collective coordinate equations (CCEs) for $K$, $c$ and $\phi$. Of course, these equations hold only when the 2DDS is formed (far from primary vessel and tumor) after an initial stage that lasts a time $t_0>0$. 

\subsection{Finding CCEs by using an approximate integral formula}
We first multiply Eq.~(\ref{eq35}) by $\partial^2P_s/\partial\xi\partial\langle\overline{F_\xi}\rangle$ (which is odd in $\xi$) and integrate over $\xi$. As the soliton decays exponentially for $|\xi|\gg 1$, it is considered to be localized on some finite interval $(-\mathcal{L}/2,\mathcal{L}/2)$.  The coefficients in the soliton formula \eqref{eq29} and the coefficients in Eq.~\eqref{eq35} depend on the slowly varying TAF concentration, therefore they are functions of $\xi$ and time and get integrated over $\xi$. The TAF varies slowly on the support of the soliton, hence we can approximate the integrals of functions $F(\xi;\psi,t)$ (varying rapidly on their first argument and slowly on their second argument) over $\xi$ by 
\begin{eqnarray}
\!\int_a^b\! F(\xi;\psi,t)\, d\xi \approx\frac{1}{\mathcal{L}}\int_a^b\!\left(\int_{-\mathcal{L}/2}^{\mathcal{L}/2} F(\xi;\psi,t)\, d\xi\!\right)\! d\psi. \label{eq37}
\end{eqnarray}
As in Eq.~\eqref{eq32}, the interval $\mathcal{I}=(a,b)$ over which we integrate should be large enough to contain most of the fully formed soliton of width $\mathcal{L}$. We have $b<L$ because the region near the tumor affects the soliton and should be excluded from the interval $\mathcal{I}$, to be specified in the next section. Similarly, $a>0$. The only odd terms in $\xi$ are the last term in the left-hand side of Eq.~\eqref{eq35} and the second to last term in Eq.~\eqref{eq36}; all other terms are even in $\xi$ and cancel out when multiplied by an odd function of $\xi$ and integrated over the interval $(-\mathcal{L}/2,\mathcal{L}/2)$. Then after integrating by parts the term proportional to $-\xi\dot{\phi}$ in Eq.~\eqref{eq35}, we obtain
\begin{eqnarray}
\dot{\phi}=\frac{2}{\beta}\frac{\int_{-\infty}^\infty\left\langle\frac{\partial\overline{F_\xi}}{\partial\xi}\left(\frac{\partial^2P_s}{\partial\xi\partial\langle\overline{F_\xi}\rangle}\right)^2\right\rangle d\xi }{\int_{-\infty}^\infty\!\left\langle\frac{\partial\overline{F_\xi}}{\partial\eta}\!\left(\frac{\partial P_s}{\partial\langle\overline{F_\xi}\rangle}\right)^2\!\right\rangle d\xi }.  \label{eq38}
\end{eqnarray}
Here, factors $1/\mathcal{L}$ in numerator and denominator cancel out and we have taken the limit as $\mathcal{L}\to\infty$ in the $\xi$-integrals with negligible error because the 2DDS decays exponentially to zero as $|\psi|\to\infty$. The brackets $\langle f(\psi)\rangle$ have been defined in Eq.~\eqref{eq32}.

 We now multiply Eq.~(\ref{eq35}) by $\partial P_s/\partial K$ (which is even in $\xi$) and integrate over $\xi$. Now all terms on the right hand side of Eq.~\eqref{eq36} produce a nonzero contribution to the integral except for the second to last one. Acting similarly, we multiply Eq.~\eqref{eq35} by $\partial P_s/\partial c$  (which is even in $\xi$) and integrate over $\xi$. From the two resulting formulas, we then find $\dot{K}$ and $\dot{c}$ as
\begin{eqnarray}
&&\dot{K}=\frac{\tilde{\mathcal{A}}_K I_{cc}-\tilde{\mathcal{A}}_c I_{Kc}}{I_{KK}I_{cc}-I_{Kc}^2},\label{eq39}\\
&&\dot{c}=\frac{\tilde{\mathcal{A}}_c I_{KK}- \tilde{\mathcal{A}}_K I_{Kc}}{I_{KK}I_{cc}-I_{Kc}^2}, \quad\label{eq40}
\end{eqnarray}
in which we have used the following definitions:
\begin{eqnarray}
I_{ij}=\int_{-\infty}^\infty\left\langle\frac{\partial P_s}{\partial i}\frac{\partial P_s}{\partial j}\right\rangle d\xi, \, i,j=K,c,\quad \label{eq41}\\ 
\mathcal{A}_j=\int_{-\infty}^\infty\left\langle\frac{\partial P_s}{\partial j}\mathcal{A}\right\rangle d\xi,\quad j=K,c,\quad \label{eq42}\\
\tilde{\mathcal{A}}_j \!=\! \mathcal{A}_j \!-\! \dot{\phi} \!\int_{-\infty}^\infty\left\langle\frac{\partial P_s}{\partial j}\frac{\partial P_s}{\partial \langle\overline{F_\xi}\rangle}\,\overline{F_\eta} \right\rangle\! d\xi, \quad j=K,c. \label{eq43}
\end{eqnarray}

\subsection{Equations for collective coordinates}
The integrals of Eqs.~\eqref{eq41}-\eqref{eq43} are calculated using Mathematica. As the coefficients $\chi$ and $\kappa$ are very small, the TAF concentration varies slowly, and terms containing them are ignored. We have also set $\mu$ to be a constant. Then Eqs.~\eqref{eq36}, \eqref{eq39}-\eqref{eq40} become
\begin{eqnarray}
\dot{K}\!&=&\! \frac{(2Kg\!+\!\langle\overline{\mu}\rangle^2)^2}{4g\beta(c\!-\!\langle\overline{F_\xi}\rangle)^2}\frac{\frac{4\pi^2}{75}\!+\!\frac{1}{5}\!+\!\!\left(\frac{2\langle\overline{F_\xi}\rangle}{5 c}\!-\!\frac{2\pi^2}{75}\!-\!\frac{9}{10}\right)\!\!\frac{\langle\overline{F_\xi}\rangle}{c}}{\left(1-\frac{4\pi^2}{15}\right)\!\left(1-\frac{\langle\overline{F_\xi}\rangle}{2c}\right)^2} \nonumber\\
\!&-&\!
\frac{2Kg+\langle\overline{\mu}\rangle^2}{g\!\left(2c-\langle\overline{F_\xi}\rangle\right)}\!\left(\dot{\phi}\langle\overline{F_\eta}\rangle\!+\!c\left\langle\frac{\partial\overline{F_\eta}}{\partial\eta}\right\rangle+\!\left\langle\overline{\mathbf{F}}\cdot\nabla_\xi\overline{F_\xi}\right\rangle\!- \frac{\langle\Delta_\xi\overline{F_\xi}\rangle}{2\beta}\!\right)\! \nonumber\\
\!&+&\!
 \frac{2Kg+\langle\overline{\mu}\rangle^2}{2g\beta(c-\langle\overline{F_\xi}\rangle)^2}\langle |\nabla_\xi\overline{F_\xi}| \rangle^2
\frac{ 1-\frac{\pi^2}{30}\!-\!\frac{3 \langle\overline{F_\xi}\rangle}{2 c} \left(1-\frac{\pi^2}{90}\right) \!+\!\frac{\langle\overline{F_\xi}\rangle^2}{2c^2}}{\left(1-\frac{\langle\overline{F_\xi}\rangle}{2c}\right)^2} , \quad  \label{eq44}
\end{eqnarray}
\begin{eqnarray}
\dot{c}&=&\!-\frac{7(2Kg+\langle\overline{\mu}\rangle^2)}{20\beta(c-\langle\overline{F_\xi}\rangle)}\frac{1-\frac{4\pi^2}{105}}{\left(1-\frac{4\pi^2}{15}\right)\!\left(1-\frac{\langle\overline{F_\xi}\rangle}{2c}\right)} 
-\frac{c}{2c-\langle\overline{F_\xi}\rangle}\!\left[c\left\langle\frac{\partial\overline{F_\xi}}{\partial\xi}\right\rangle\! \right. \nonumber\\
&+&(c\!-\!\langle\overline{F_\xi}\rangle)\!\langle\nabla_\xi\!\cdot\!\overline{{\mathbf F}}\rangle\!+\!\left.\frac{\langle\Delta_\xi\overline{F_\xi}\rangle}{2\beta}-\!\left\langle\overline{\mathbf{F}}\!\cdot\!\nabla_\xi\overline{F_\xi}\right\rangle-\dot{\phi}\langle\overline{F_\eta}\rangle\right]\! \nonumber\\
&-&\frac{ \langle |\nabla_\xi\overline{F_\xi} | \rangle^2\!\left( 1+\frac{\pi^2}{30} \right) }{\beta \left( c\!-\!\langle\overline{F_\xi}\rangle\right)\! \left( 2-\frac{\langle\overline{F_\xi}\rangle}{c} \right) }. \quad  \label{eq45}
\end{eqnarray}
If we set $\dot{\phi}=0$, these equations become Eqs.~(C12)-(C13) of Ref.~\cite{bon16pre} with $\mu_C=0$ and $\xi=x$. The coefficients entering Eqs.~\eqref{eq44} and \eqref{eq45} are spatial averages over $\psi$ (which is the slow variable $\xi$ that appears in the formulas through the TAF concentration) and have $\eta=0$ due to Eq.~\eqref{eq22a}. The CCEs \eqref{eq38}, \eqref{eq44} and \eqref{eq45}, describe the mean behavior of the 2DDS after its formation time, whenever it is far from primary vessel and tumor, as we will show in the next section. 

\section{Numerical results}
\label{sec:numerical}
In this paper, we obtain the vessel tip density by ensemble averages of stochastic simulations, as explained in Ref.~\cite{ter16}. If we set up symmetric initial and boundary conditions so that the 2DDS moves on the $x$-axis, $\mathbf{X}=(X,0)$, $\hat{\mathbf{V}}=(1,0)$, $\hat{\mathbf{V}}^\perp= (0,1)$, $\xi=x-X$, $c=\dot{X}$, and $\eta=y$. Then the integrals in Eq.~\eqref{eq23} are integrals over $y$ and the integrals in Eq.~\eqref{eq32} are simply integrals over $x$ with $y=0$. From our simulations, we can obtain the evolution of the 2DDS collective coordinates thereby reconstructing the marginal tip density from Eqs.~\eqref{eq22}, \eqref{eq29}, and \eqref{eq44}-\eqref{eq45} with $\dot{\phi}=0$. The 2DDS profile at $y=0$ is the 1D soliton studied in Refs.~\cite{bon16,bon16pre}, which agrees with numerical simulations of the stochastic process and also with simulations of the corresponding deterministic equations. 

\subsection{Initial and boundary conditions}
The simplest asymmetric configuration consists of one initial tip moving toward a TAF source at $x=1$, above the $x$-axis. To the values of the dimensionless parameters in Table \ref{table2}, we have added the initial nondimensional TAF concentration
		\begin{eqnarray}
		C(0,x,y) = 1.1\, e^{-(x-1)^2/1.5^2 -(y-0.4)^2/0.6^2}, \label{eq46}
		\end{eqnarray}
and the nondimensional TAF flux boundary condition at $x = 1$
		\begin{eqnarray}
		\frac{\partial C}{\partial x}(t,1,y) = 1.1 e^{-(y-0.4)^2/0.6^2}. \label{eq47}
		\end{eqnarray}
At $x=0$, the TAF flux is zero. These conditions correspond to having a TAF source at the border $x=1$, above the $x$-axis at  $y=0.4$. 

\begin{figure}[h]
\begin{center}
\includegraphics[width=14cm]{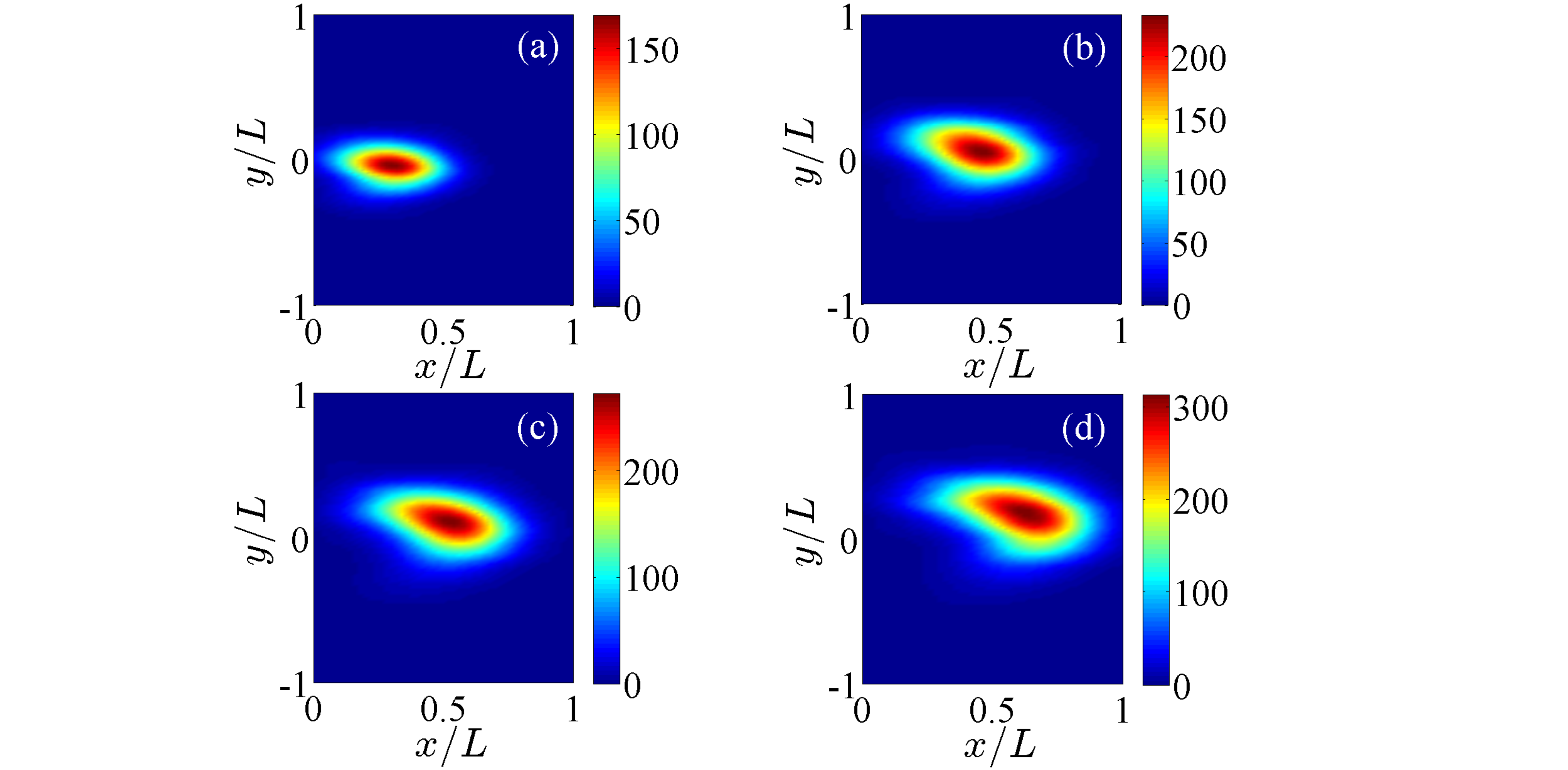}
\end{center}
\vskip-3mm 
\caption{Density plots of the marginal tip density $\tilde{p}(t,x,y)$ calculated from Eq.~\eqref{eq8} with $N(0)=1$ and $\mathcal{N} = 400$ replicas, showing how tips are created at $x = 0$ and march toward the tumor at $x = L$. Snapshots at (a) 16 hr, (b) 24 hr, (c) 28 hr, (d) 32 hr.} \label{fig9}
\end{figure}

\begin{figure}[h]
\begin{center}
\includegraphics[width=14cm]{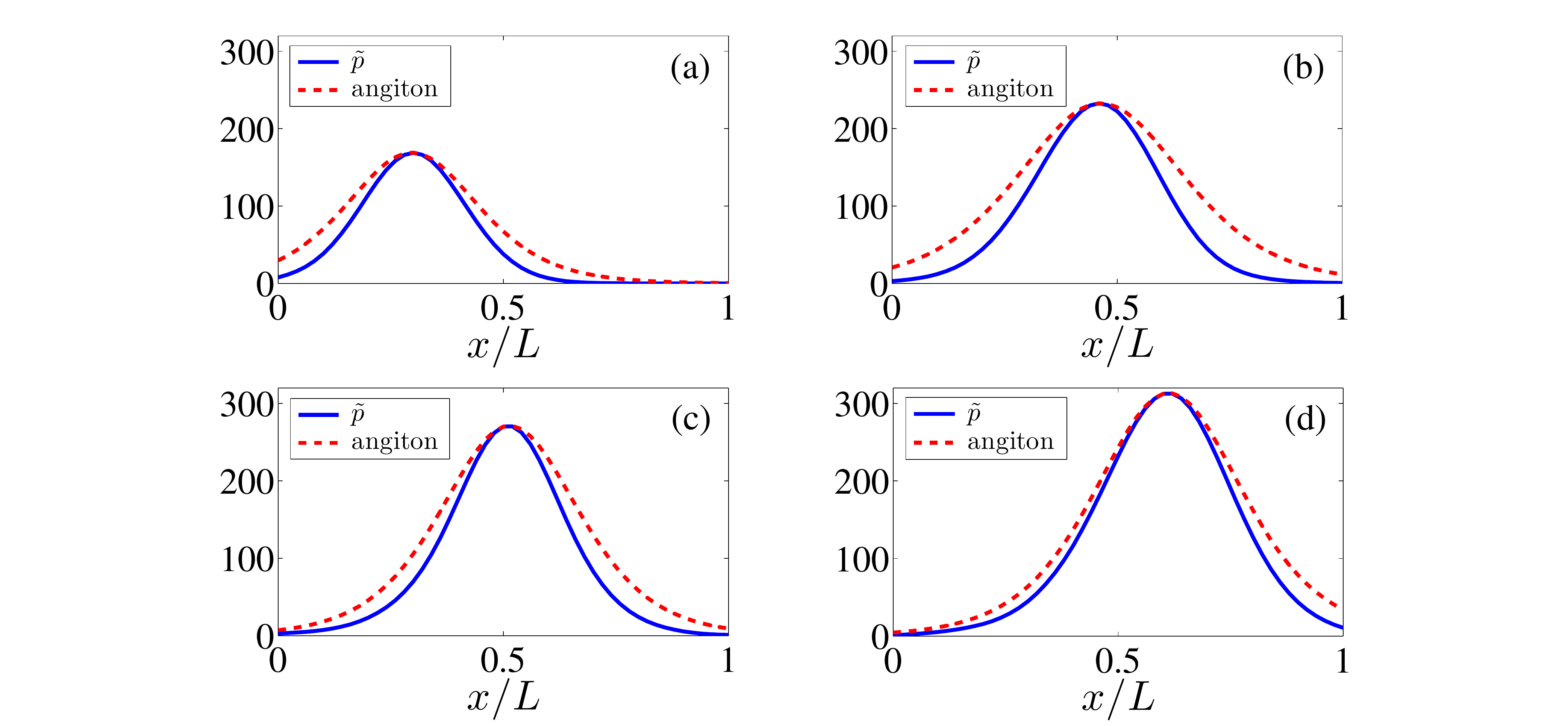}
\end{center}
\vskip-3mm 
\caption{Profiles of the marginal tip density $\tilde{p}(t,\mathbf{x})$ for $\eta=0$  calculated as in Fig.~\ref{fig9}, and for the same times. We have fit the 2DDS (the angiton) given by Eqs.~\eqref{eq22a} and \eqref{eq29} with $\overline{F}_\xi$ calculated at $t=16$ hr and fixed for later times. For each time, $c$ and $K$ are calculated from the maximum of $\tilde{p}(t,\mathbf{x})$ and the trajectory of its center of mass.} \label{fig10}
\end{figure}

\subsection{A single initial tip}
Suppose that initially there is only one tip, $N(0)=1$, placed below the $x$-axis, say at $(0,-0.2)$. In a typical realization of the stochastic process, the initial active tip advances and undergoes repeated branching until the density of active tips approaches the 2DDS. For sufficiently large distance between the primary blood vessel and the tumor, the evolution of the active tip density comprises three stages: a soliton formation stage, evolution of the 2DDS far from the boundaries, and arrival at the TAF source. Here, we only describe the second stage of a 2DDS detached from the boundaries. A complete theory would require matching the detached soliton stage to reduced descriptions of the other stages, which we do not attempt in this paper. For shorter distances between primary blood vessel and TAF source, a 2DDS may not even form and our theory is then inapplicable. 

The  evolution of the angiogenic network and the duration of the 2DDS formation period depend on the specific selection of the velocity in Eq.~\eqref{eq6}. For example, if $\mathbf{v}_0$ is parallel to the $x$ axis, it takes 18 hr to form the 2DDS, which finds it difficult to move upward to where the TAF source of Eq.~\eqref{eq47} is. Furthermore, there are more than 30 realizations of the stochastic process for which anastomosis eliminates all active tips before they reach $x=1$. We need to discard these replicas when calculating the density of active tips by an ensemble average. A better choice of $\mathbf{v}_0$ decreases the number of replicas to be discarded and lifts the center of mass for the angiogenic network of active tips. Fig.~\ref{fig9} shows four snapshots of $\tilde{p}(t,\mathbf{x})$ after the 2DDS formation time for $\mathbf{v}_0=(1,0.4)$ and a 400-replica ensemble average. For this modified $\mathbf{v}_0$, only two replicas need to be discarded. From the trajectory of the maximum of $\tilde{p}$, which coincides with the 2DDS center of mass, we calculate $K$, $c$ and $\phi$. Fig.~\ref{fig10} displays a comparison of the snapshots of Fig.~\ref{fig9} for $\eta=0$ with the 2DDS obtained from Eqs.~\eqref{eq22a} and \eqref{eq29}. The reasonably accurate fit shown in Fig.~\ref{fig10} confirms the validity of the 2DDS description after the formation stage and before the arrival at the tumor. 

\subsection{Coefficients in the CCEs and initial conditions after the 2DDS formation stage}
As explained before, when there is a single tip at $t=0$, the 2DDS formation stage takes longer, certain realizations of the stochastic process end up prematurely by anastomosis before the tips can reach $x=1$ and have to be discarded. In addition, the influence of the details of tip velocity selection at branching disappears in the overdamped limit. However, these different details still affect the 2DDS motion. Thus, we shall compare its motion to an initial configuration that has a faster formation stage and does not require discarding failed replicas of the stochastic process. Let us now consider an asymmetric configuration as in Fig.~\ref{fig1}. In each stochastic simulation (replica), $N(0)=20$ initial tips are placed at $x=0$ and uniformly distributed in the $y$-direction between -0.5 and 0.1 with $\mathbf{v}_0=(1,0)$. 

Stochastic simulations indicate that it takes a time $t_0=0.318$ (16 hours) after angiogenesis initiation to form the 2DDS. For $t>t_0$, the evolution of the soliton is given by Eqs.~\eqref{eq44}-\eqref{eq45}. The variance in Eq.~\eqref{eq22} is fixed as $\sigma=0.235$. As indicated before, we consider that the collective coordinates represent spatial averages over the spatial coordinate $x$ excluding regions affected by boundaries. The coefficients in Eqs.~\eqref{eq44}-\eqref{eq45} are spatial averages involving $C(t_0,x,y)$. We calculate them by: (i) approximating all differentials by second order finite differences, (ii) approximating the integrals in Eq.~\eqref{eq23} by Gaussian quadrature, and (iii) using Eq.~\eqref{eq32} to average the coefficients by taking the arithmetic mean of their values at all grid points in the interval $x\in\mathcal{I}=(0.54,0.95]$ ($0.21<\xi\leq 0.63$). For $0<x\leq 0.54$ and for $0.95<x\leq 1$, the boundary conditions at $x=0$ and at $x=1$, respectively, influence the outcome and therefore we leave these values out of the averaging. 

\begin{figure}[h]
\begin{center}
\includegraphics[width=6cm]{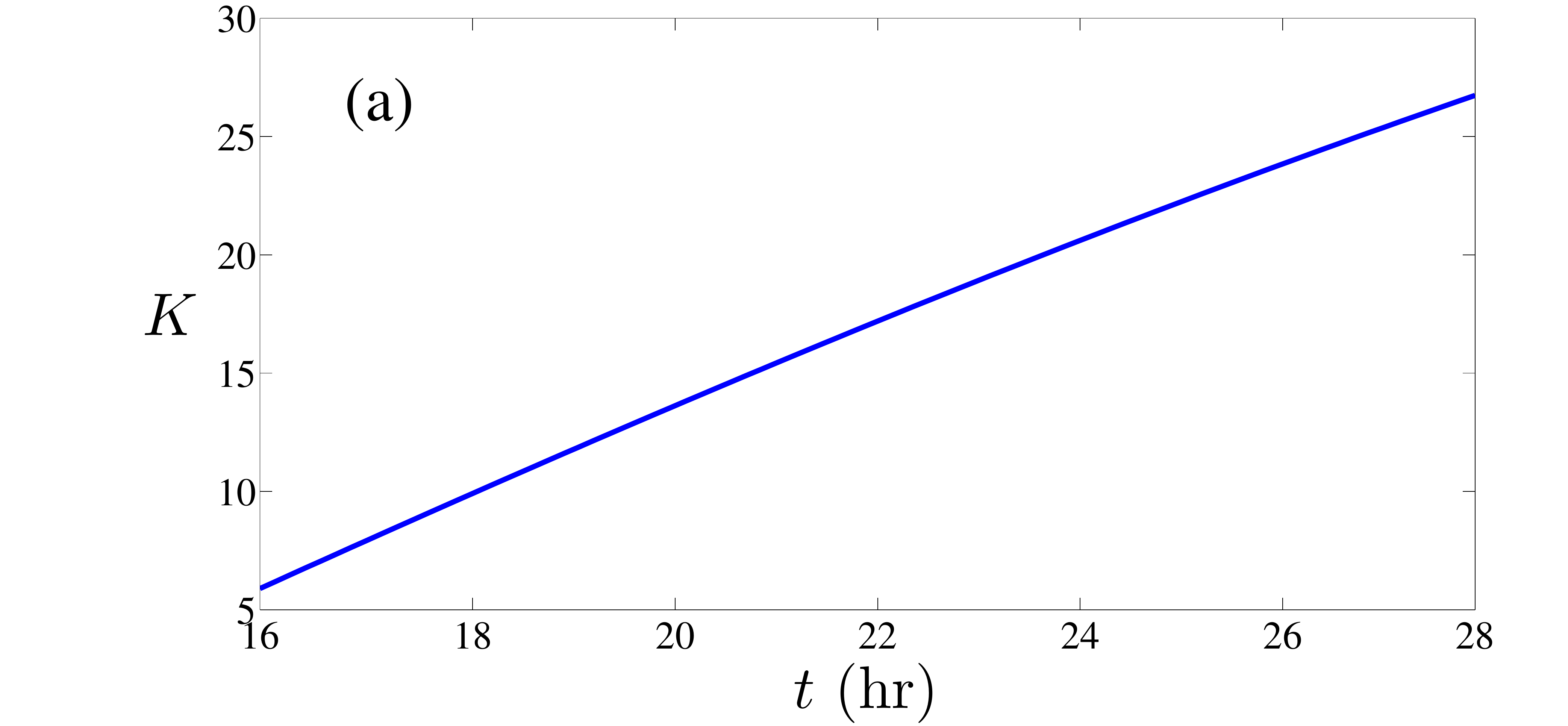} 
\includegraphics[width=6cm]{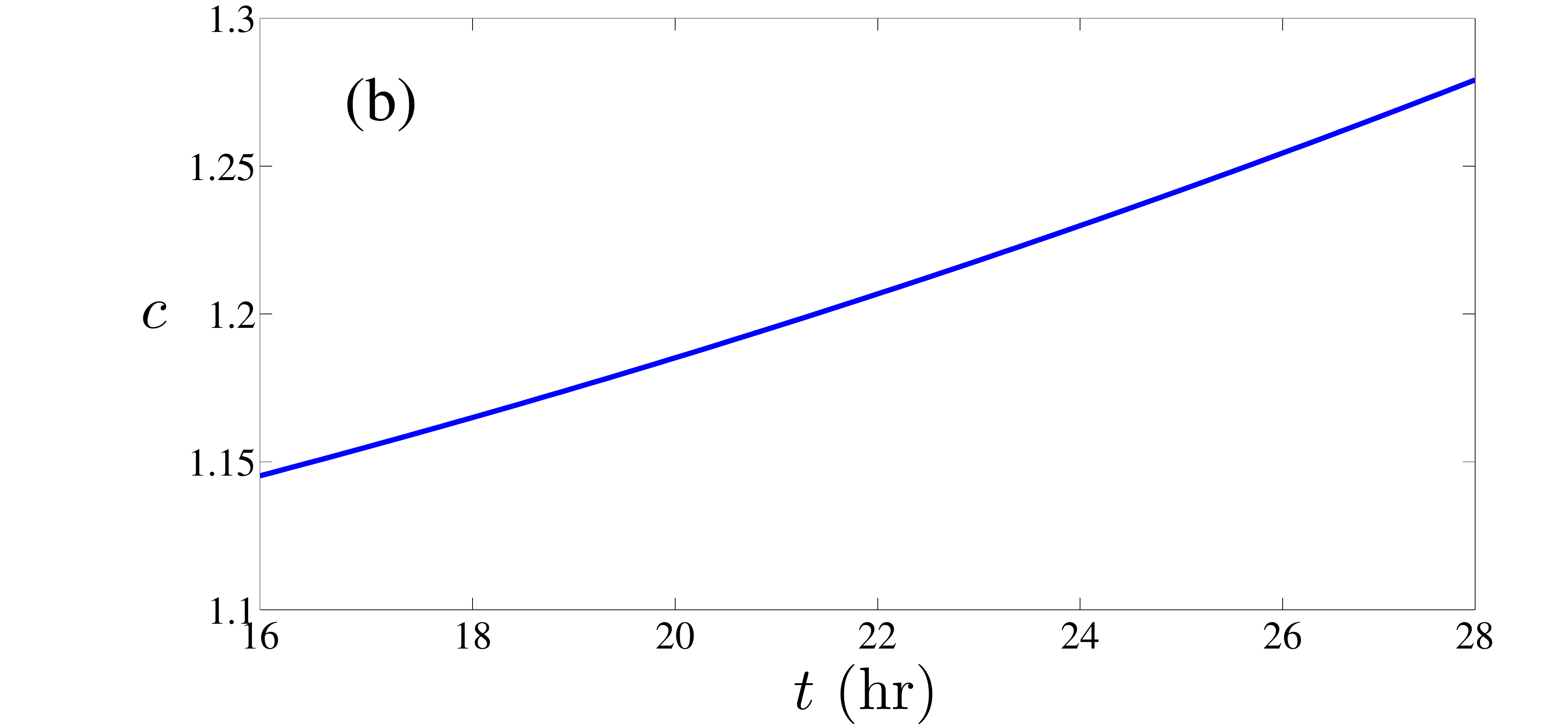} 
\includegraphics[width=6cm]{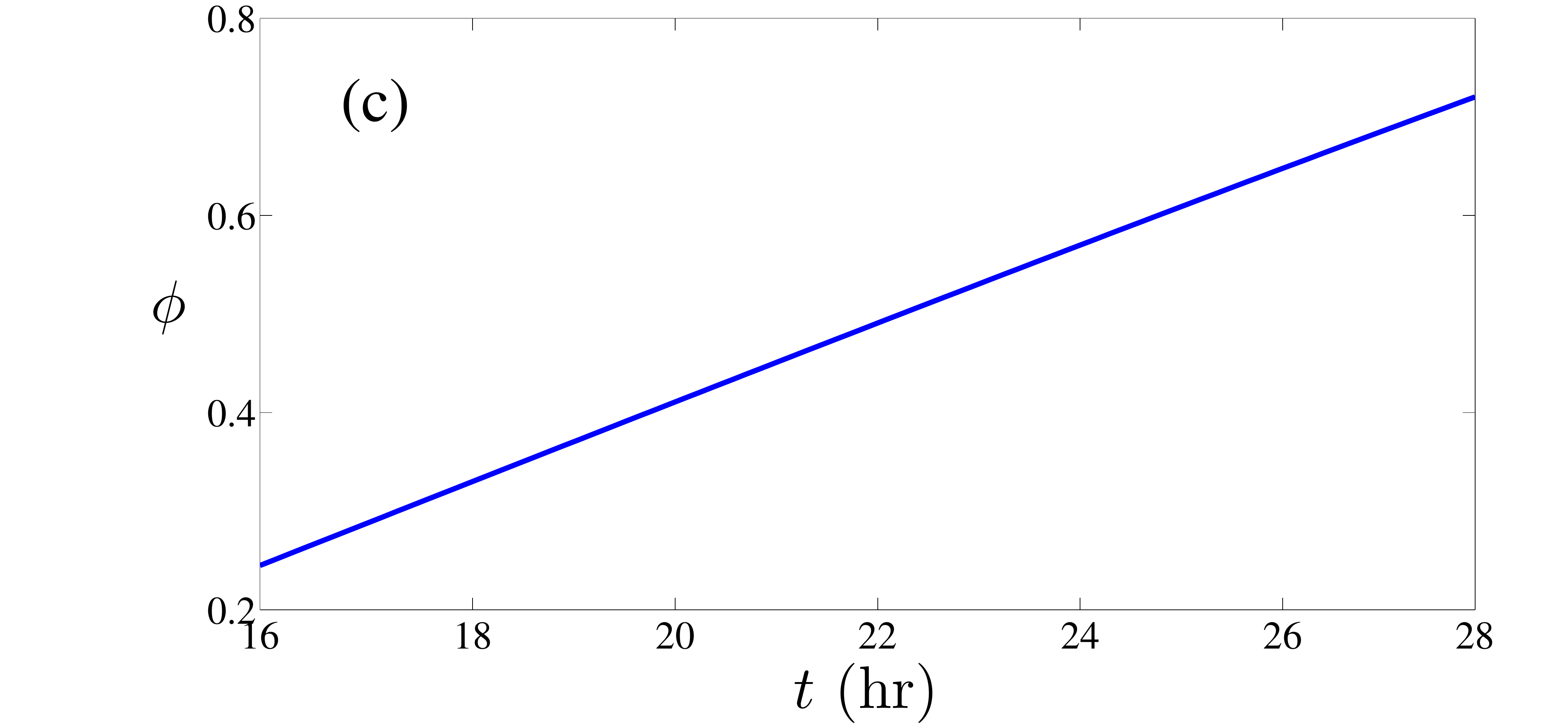}
\includegraphics[width=6cm]{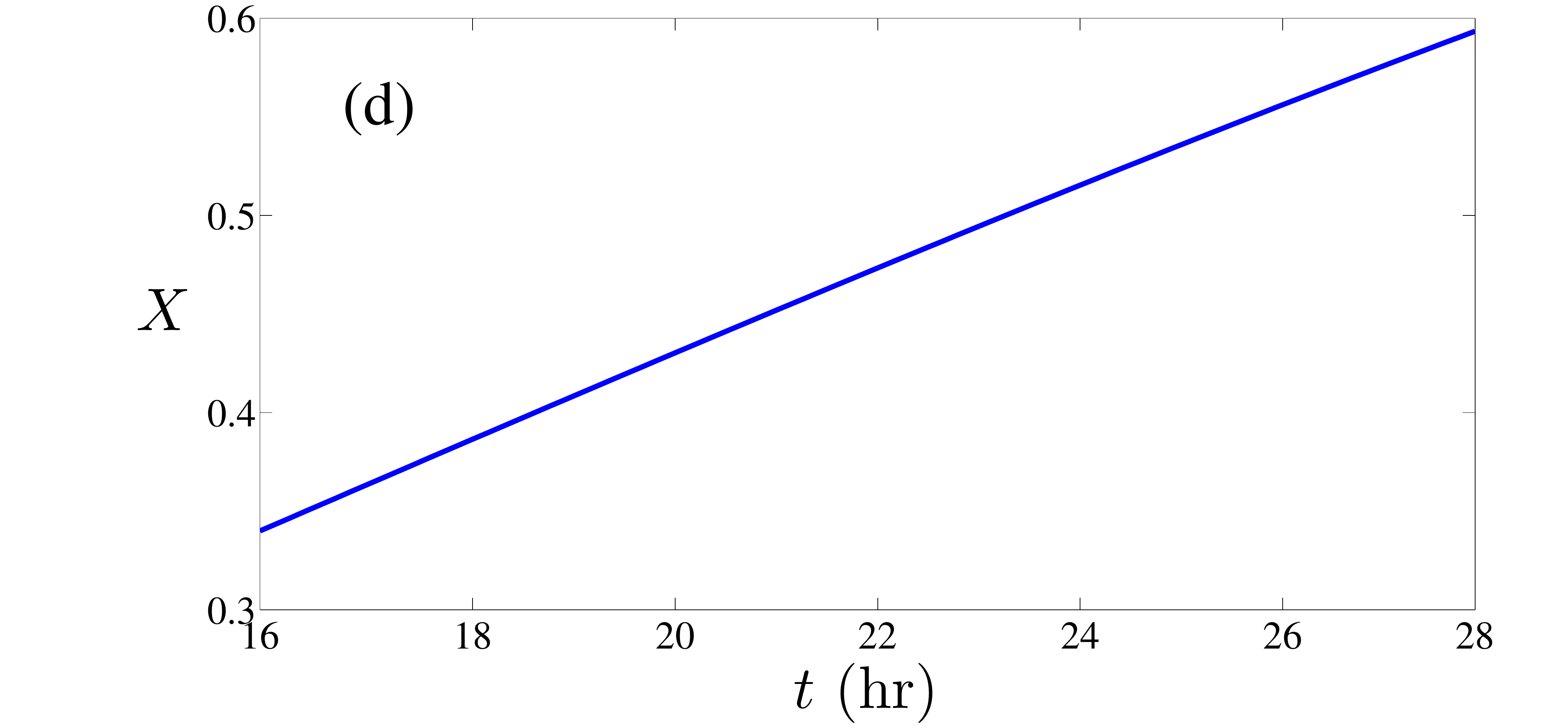}
\includegraphics[width=6cm]{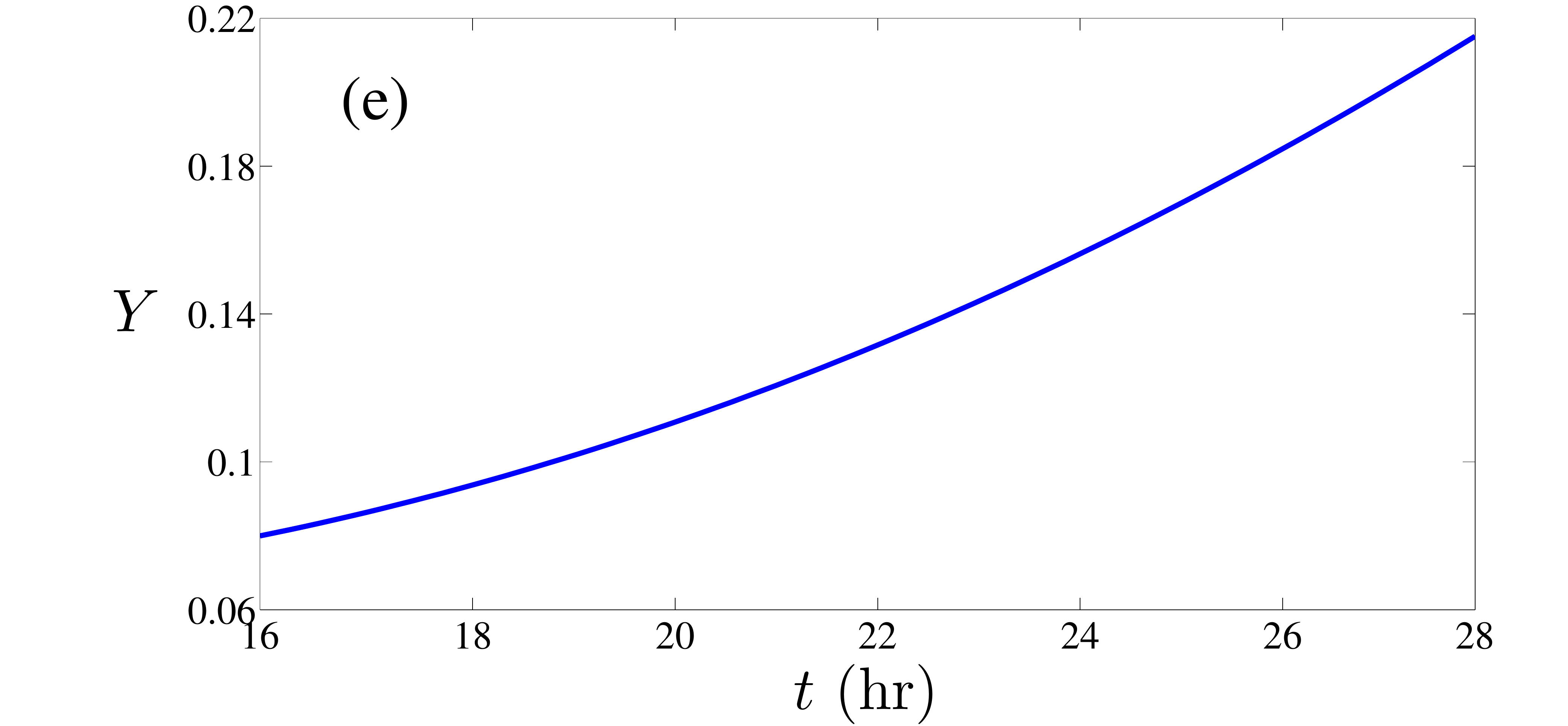}
\end{center}
\vskip-3mm 
\caption{ Evolution of the collective coordinates (a) $K(t)$, (b) $c(t)$, (c) $\phi(t)$, (d) $X(t)$, (e) $Y(t)$. \label{fig3}}
\end{figure}

The initial conditions for the CCEs \eqref{eq44}-\eqref{eq45} are set as follows. We find the coordinates of the maximum of the marginal tip density $\tilde{p}(t_0,x,y)$ (calculated from ensemble average by Eq.~\eqref{eq8} with $\mathcal{N}=400$) as $\mathbf{X}(t_0)=\mathbf{X}_0 = (0.34,0.08)$. Similarly, we set $K(t_0)= 5.9$, $c(t_0)=1.15$, $\phi(t_0)=0.245$, determined so that the maximum marginal tip density at $t=t_0$ coincides with the soliton peak. Solving the CCEs \eqref{eq44}-\eqref{eq45} with these initial conditions, we obtain the curves depicted in Fig.~\ref{fig3}.

\begin{figure}[h]
\begin{center}
\includegraphics[width=7cm]{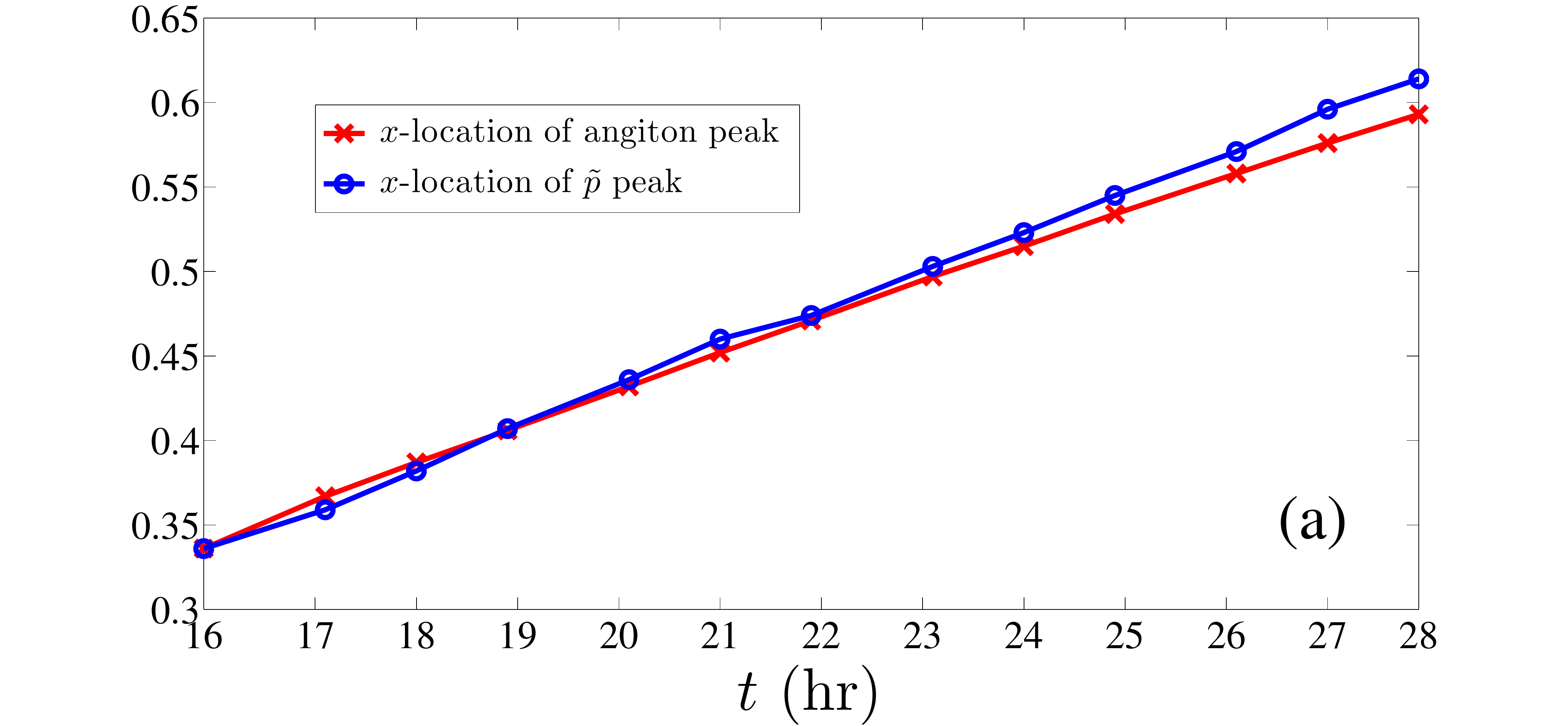}
\includegraphics[width=7cm]{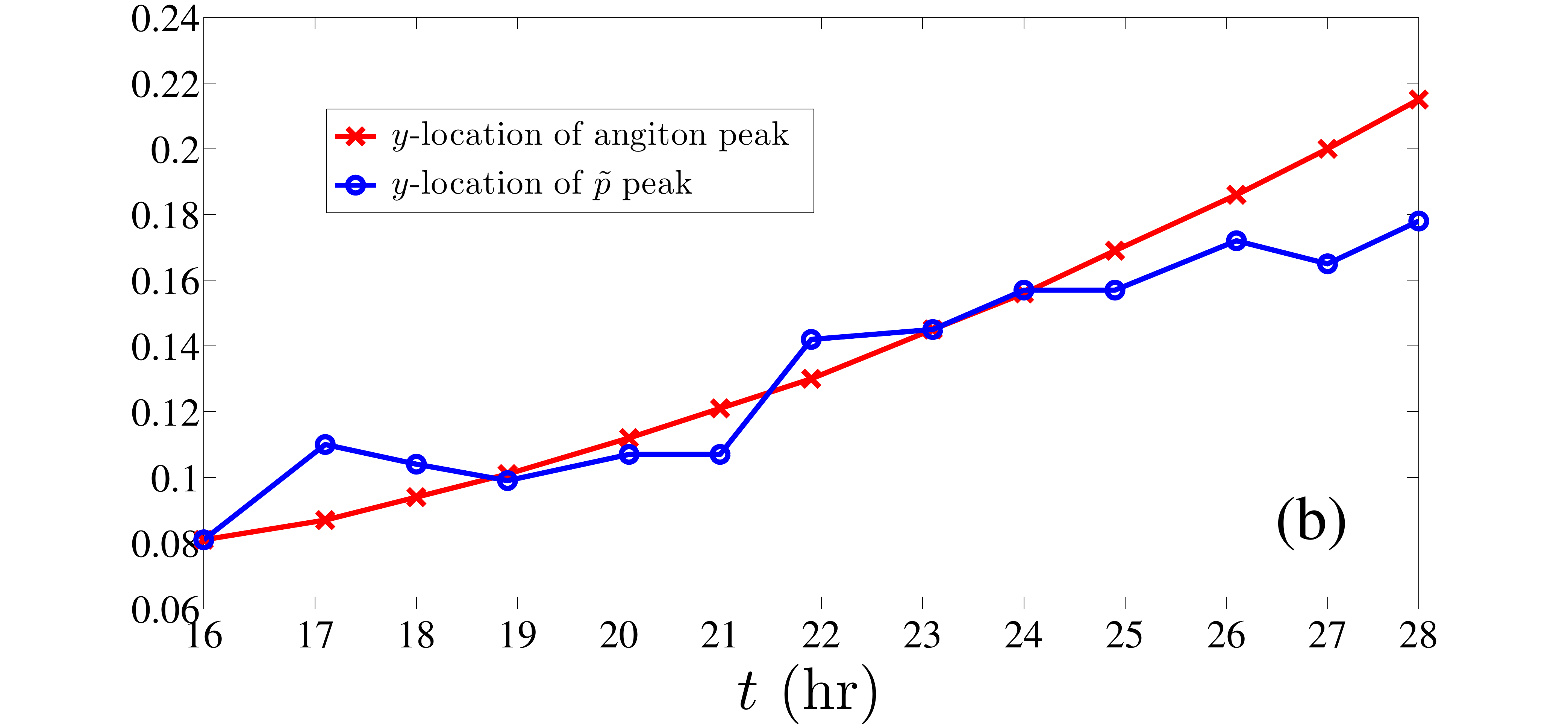}
\includegraphics[width=7cm]{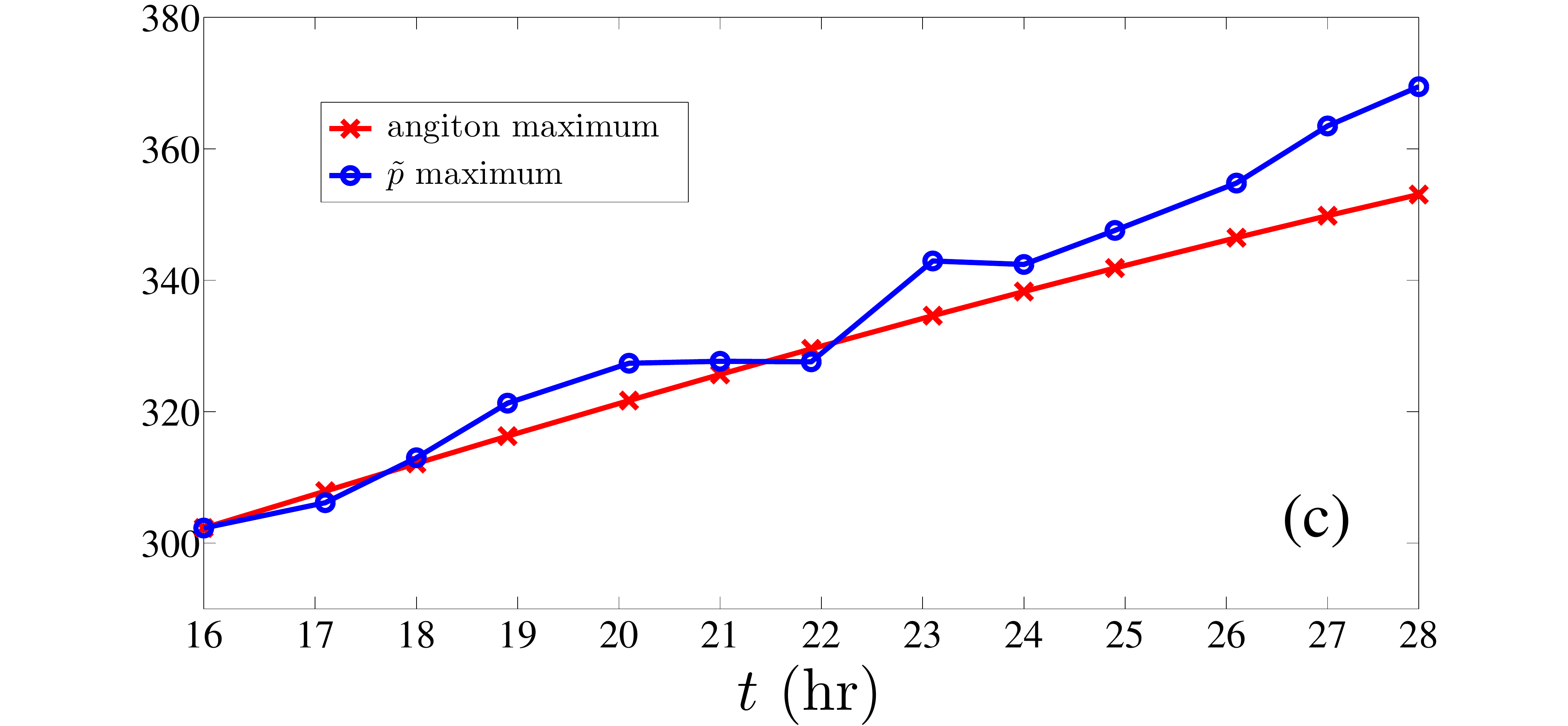} 
\end{center}
\vskip-4mm \caption{Evolution of (a) $x$-coordinate, (b) $y$-coordinate, and (c) value of the ensemble-averaged maximum marginal tip density (by $\mathcal{N} = 400$ replicas) as compared to the prediction by the collective coordinates of the 2DDS (the {\em angiton}). The maximum absolute error is reached at $t = 27$ hr, $28$ hr. It approximately equals (a) $\Delta x = 0.02$ and (b) $2\Delta x = 0.04$, where $\Delta x$ is the discretization space step. In (c) the relative error is around 4\% at $t = 27$ hr, $28$ hr, while it is smaller than 2.5\% for all other times. \label{fig4}}
\end{figure}
\begin{figure}[htbp]
\begin{center}
\includegraphics[width=14.0cm]{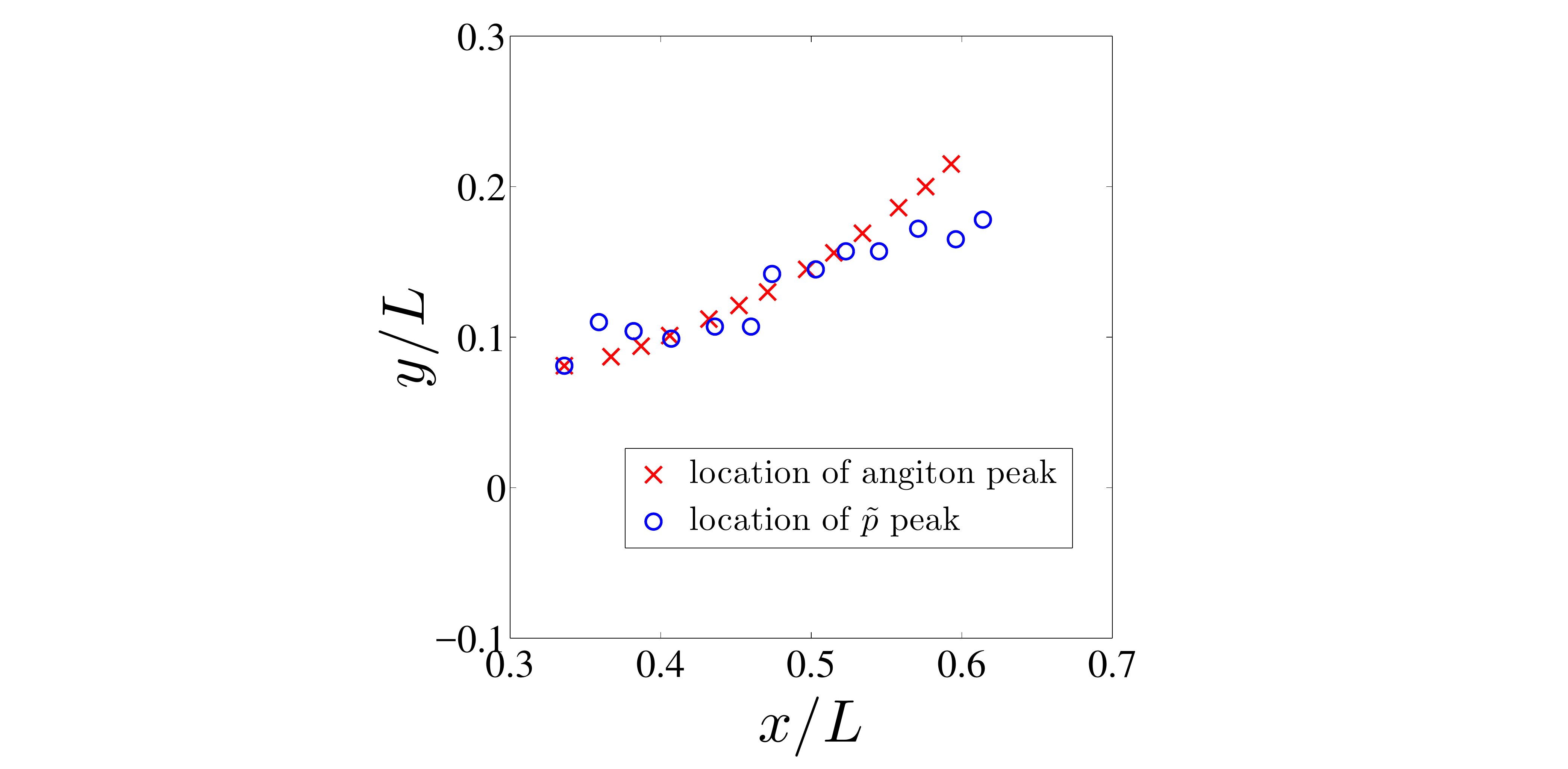}
\end{center}
\vskip-4mm \caption{Evolution of the position $(x,y)$ of the ensemble-averaged maximum marginal tip density (over $\mathcal{N}=400$ replicas) as compared to that of the 2DDS ({\em angiton}) predicted by collective coordinates. The absolute error is approximately equal to $2\Delta x = 0.04$ at $t = 27$ hr, $28$ hr, while it is $\Delta x = 0.02$ at most for all other times. Here $\Delta x$ is the space step in the discretization. \label{fig5}}
\end{figure}

\subsection{Comparison of CCE predictions with stochastic simulations}
Using the 2DDS collective coordinates depicted in Fig.~\ref{fig3} and Eqs.~\eqref{eq22} and (\ref{eq29}), we reconstruct the marginal vessel tip density and find its maximum value and the location thereof for all times $t>t_0$. Fig.~\ref{fig4} shows that the position of the  2DDS as predicted from the CCEs \eqref{eq38} and \eqref{eq44}-\eqref{eq45} compares very well with the location of the tip density maximum obtained by ensemble average of stochastic simulations (over $\mathcal{N}=400$ replicas). Fig.~\ref{fig5} shows that the overall trajectory of the 2DDS agrees well with the location of the tip density maximum, which coincides with the 2DDS center of mass. 

\begin{figure}[htbp]
\begin{center}
\end{center} 
\includegraphics[width=14cm]{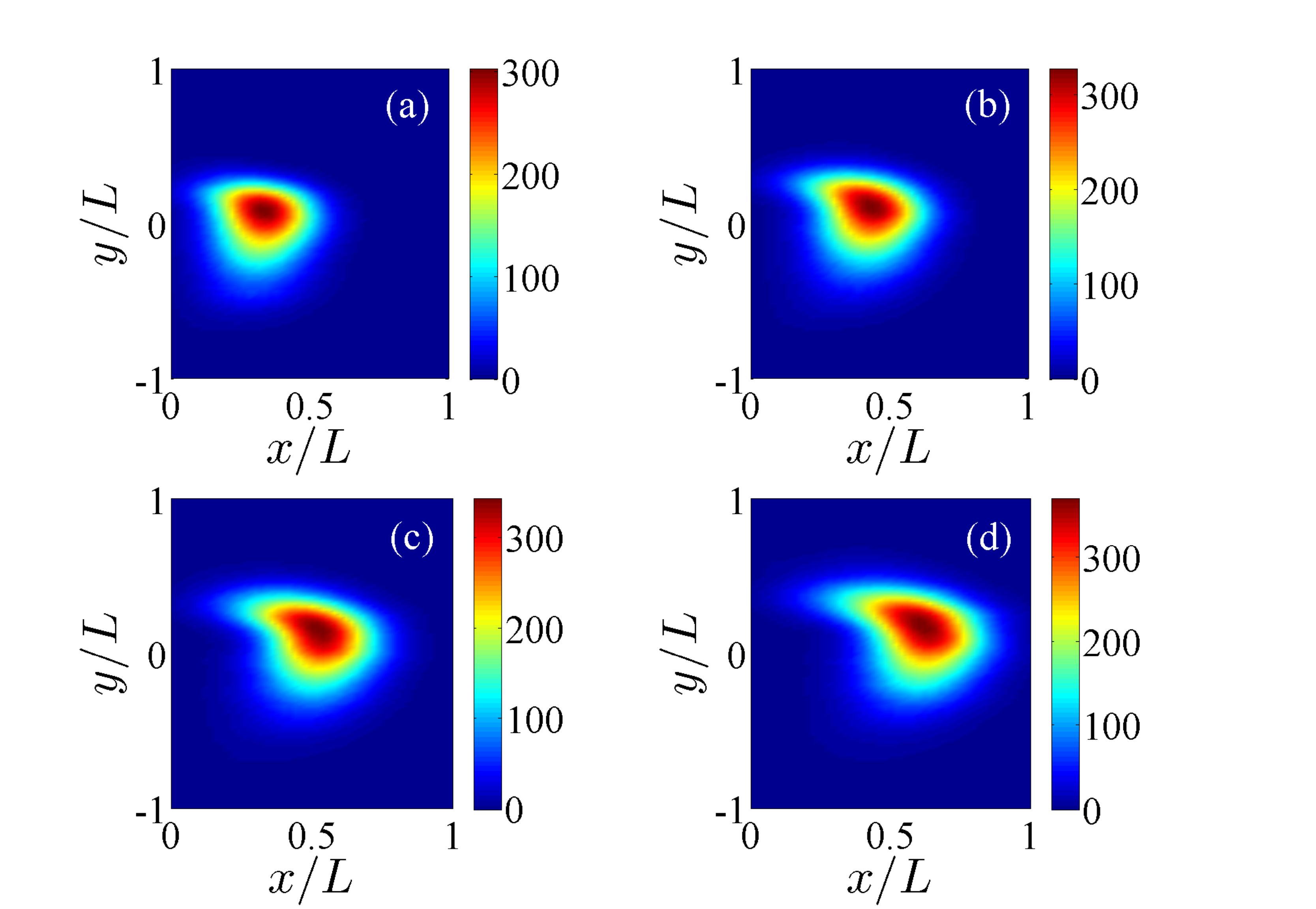}\\ 
\caption{Density plots of the marginal tip density $\tilde{p}(t,x,y)$ calculated from Eq.~\eqref{eq8} with $\mathcal{N} = 400$ replicas, showing how tips are created at $x = 0$ and march toward the tumor at $x = L$. Snapshots at (a) 16 hr, (b) 20 hr, (c) 24 hr, and (d) 28 hr.
\label{fig6}}
\end{figure}
\begin{figure}[htbp]
\begin{center}
\end{center} 
\includegraphics[width=16cm]{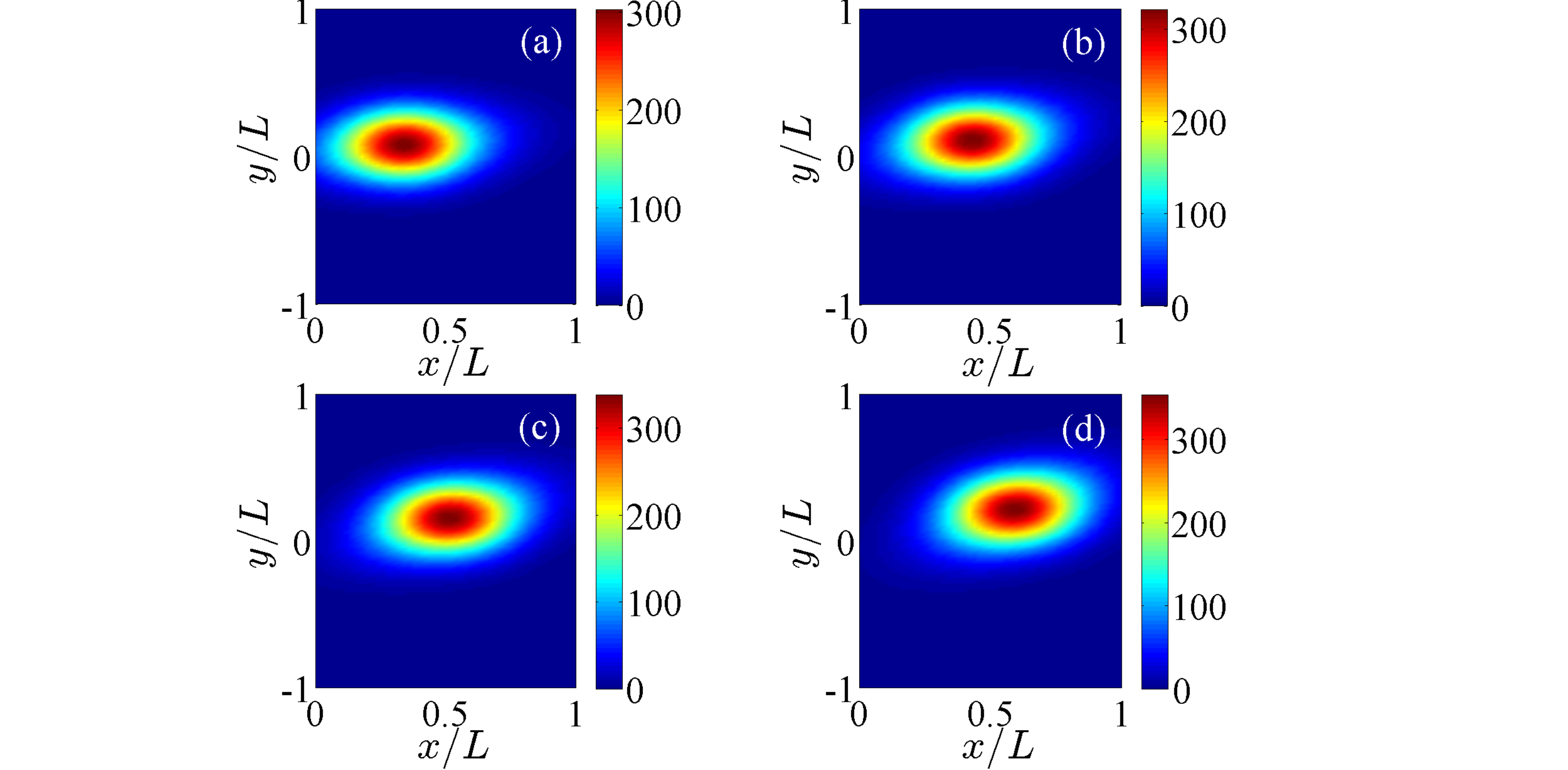}
\caption{Same as in Fig.~\ref{fig6} but now the density plots of $\tilde{p}(t,x,y)$ are calculated
from Eq.~\eqref{eq22} and the CCEs \eqref{eq38}, \eqref{eq44}, \eqref{eq45} for the 2DDS. Snapshots at (a) 16 hr, (b) 20 hr, (c) 24 hr, and (d) 28 hr.
\label{fig7}}
\end{figure}

Fig.~\ref{fig6} is the density plot of the ensemble-averaged marginal tip density in four snapshots taken at 16, 20, 24 and 28 hours after the initial time. Fig.~\ref{fig7} shows the density plot of the marginal tip density  reconstructed from the 2DDS CCEs. The shape of the respective density plots is different but the sizes of their peaks are similar, which suggests correcting the leading order of the multiple scales theory, Eq.~\eqref{eq22}. However, Figs.~\ref{fig4} and \ref{fig5} show that the 2DDS center of mass gives a good approximation for the motion of the marginal density peak. Thus, the 2DDS gives a good approximation of the advance of the marginal density and its order of magnitude. This agreement is all the more remarkable, as the parameters in Table \ref{table2} used in our simulations are not particularly small. 

\begin{figure}[h]
\begin{center}
\begin{tabular}{c}
\includegraphics[width=12.0cm]{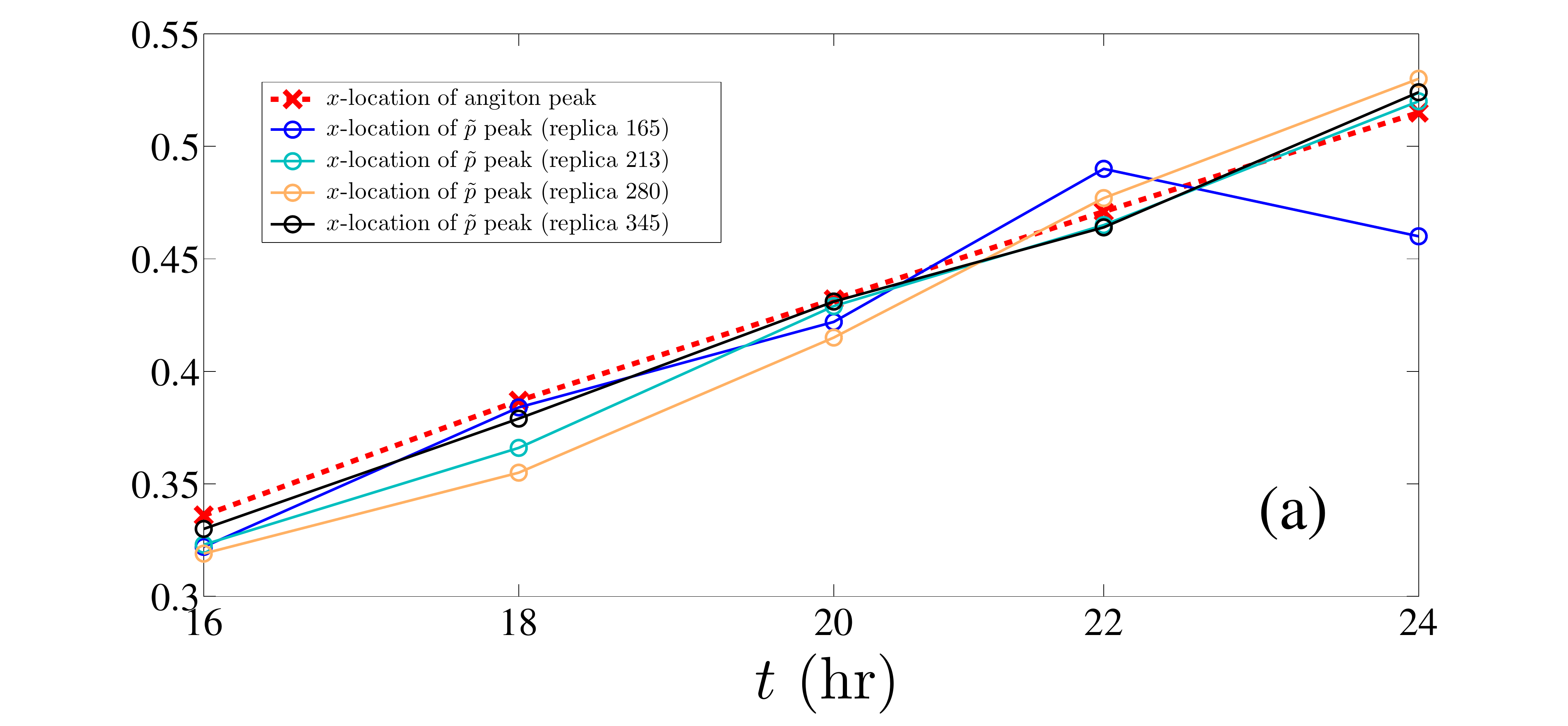}\\
\includegraphics[width=12.0cm]{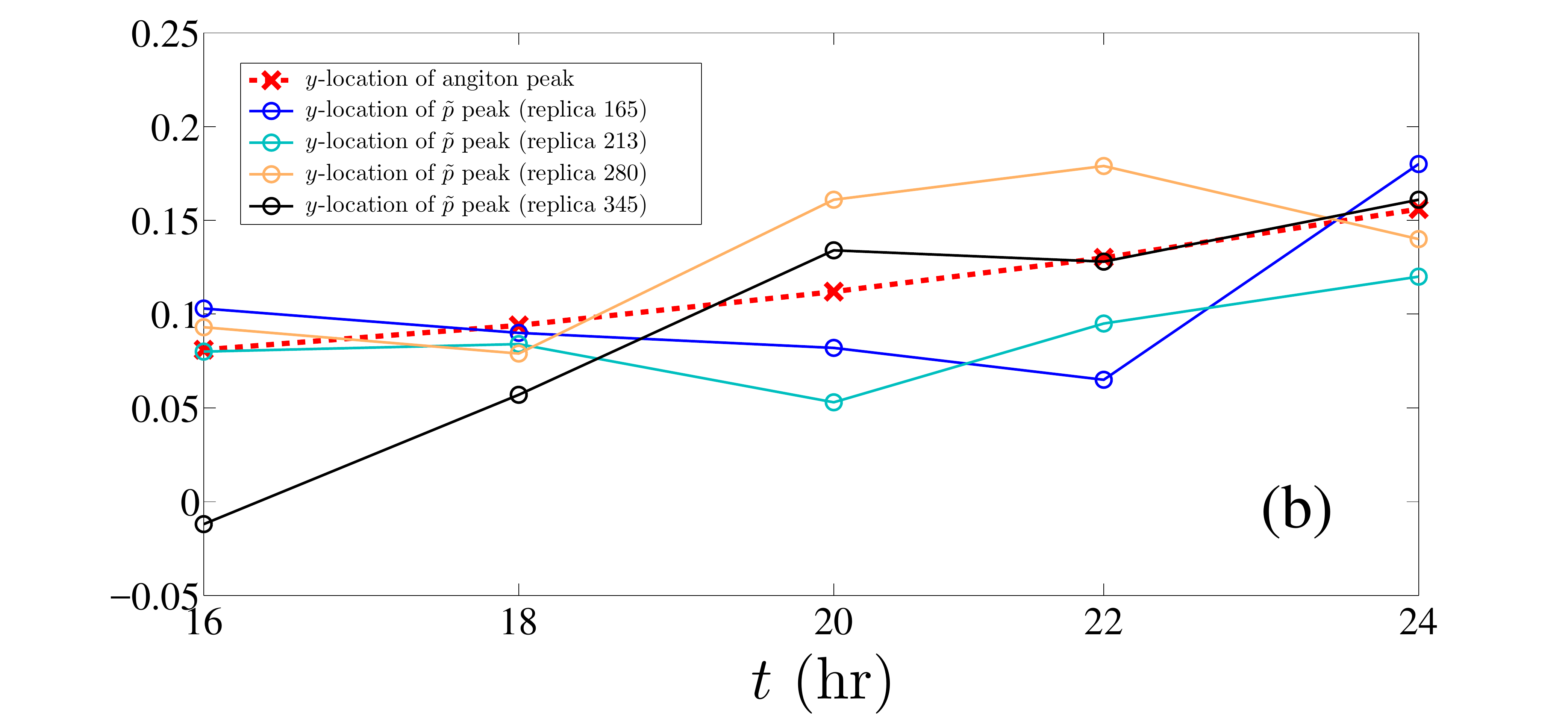}
\end{tabular}
\end{center}
\vskip-4mm \caption{(a) $x$-coordinate and (b) $y$-coordinate of the 2DDS (angiton) peak compared to those of the maximum marginal tip density for different replicas of the stochastic process. \label{fig8}}
\end{figure}

\subsection{2DDS center of mass and replicas of the stochastic process}
So far, our reconstructions have been based on ensemble averages, which produce mean values of the density of active tips and related functions. In past work, we have shown that fluctuations about the mean are large and, therefore, the stochastic process is not self-averaging \cite{ter16}. However, Fig.~\ref{fig8} indicates that the 2DDS center of mass is a good approximation to the location of the maximum marginal tip density for different replicas of the stochastic process. The $x$-coordinate of the maximum density location is approximated better than its $y$-coordinate. While vessel networks may widely differ from replica to replica, the position of the maximum marginal tip density is about the same for different replicas. As the maximum of the marginal tip density is a good measure of the advancing vessel network, the location of the 2DDS peak also characterizes it. 

\section{Conclusions}
\label{sec:conclusions}
On mesoscopic distances that are large compared to the size of one cell but small compared to the size of an organ, the early stage of tumor induced angiogenesis can be described by stochastic models that track the trajectories of active vessel tips. These models consider branching of blood capillaries as a stochastic process and renounce to describe cellular processes and scales. However, active tip models pose novel and interesting problems in nonequilibrium statistical mechanics. In previous works, we have shown that the ensemble-averaged density of active tips is described by an integrodifferential Fokker-Planck equation with source and sink terms \cite{bon14,ter16}. Together with time derivative and linear convection, these terms make it possible for this equation to have an approximate soliton solution for simple one dimensional geometries \cite{bon16,bon16pre,bon17}. The soliton solution has the same shape as the well-known Korteweg-de Vries soliton, which results from a balance between time derivative, nonlinear convection and dispersion \cite{lax68}. 

In two dimensions, the marginal density of active tips acquires the form of a moving lump or 2DDS that advances towards the tumor in a curvilinear trajectory. Here we have used a method of multiple scales to show that the transversal section of the 2DDS is a narrow Gaussian and that its longitudinal section is a diffusive soliton. The slow variation of the tumor angiogenic factor changes slowly the shape and trajectory of the 2DDS. The latter can be reconstructed by solving collective coordinate equations for its speed, direction of velocity, shape parameter and coordinates of the center of mass. As the parameters used in numerical simulations are not particularly small, it is remarkable that the predictions from the 2DDS and its collective coordinates (based on the method of multiple scales and singular perturbation ideas) compare well with the predictions from numerical simulations of the stochastic model. The shape of the marginal density of active tips as obtained from ensemble averages of stochastic simulations is less symmetric than that reconstructed from the 2DDS. However, the size and position of its peak follow those given by Eqs.~\eqref{eq22a} and \eqref{eq29} (the 2DDS) and the CCEs \eqref{eq38}, \eqref{eq44} and \eqref{eq45}.

In principle, the present model and the 2DDS construction can be extended to three spatial dimensions. We need to consider the angiogenic vessels as the tip trajectories plus a narrow region or tube around them and a criterion for anastomosis; cf Ref.~\cite{cap19} for a possible way to do this. The derivation of CCEs could proceed along the lines explained in the present work. We would need to consider an additional curvilinear coordinate along the binormal and modify accordingly the CCEs. Extensions of the present work to more complete models based on tip cell motion are possible \cite{cap09,hec15,bon17}, but we feel it is better to present these ideas in the simplest possible context. Future applications of the present work to biology include investigating possible control of the 2DDS motion, e.g. by studying the effect of antiangiogenic drugs; cf Ref.~\cite{lev01}. In physics, our work could be useful to study dynamic phenomena that include stochastic branching and merging of advancing point defects. For example, propagation of cracks in brittle materials \cite{ala10} or dielectric breakdown \cite{bea88}. From the point of view of nonequilibrium statistical mechanics, at the present time there is no theory of the large fluctuations about the averaged tip density. Perhaps deriving functional equations for the moments and using ideas similar to those appearing in turbulence theory could be helpful \cite{bir13}. 

\section*{Acknowledgements}
This work has been supported by the FEDER/Ministerio de Ciencia, Innovaci\'on y Universidades -- Agencia Estatal de Investigaci\'on grant MTM2017-84446-C2-2-R.

\appendix
\setcounter{equation}{0}
\renewcommand{\theequation}{A.\arabic{equation}}
\section{Boundary and initial conditions for the deterministic equations}\label{app1}
The governing equations \eqref{eq10} and \eqref{eq12} of the deterministic description have to be solved with appropriate initial and boundary conditions compatible with the stochastic description. The nondimensional initial and boundary conditions for the TAF are Eqs.~\eqref{eq46} and \eqref{eq47}, respectively, and we also have $\lim_{y\to\pm\infty} C= 0$  \cite{bon16pre}. We do not intend to follow the process of angiogenesis beyond the time when vessel tips have arrived at the tumor and therefore we do not give the latter a finite length. The boundary conditions for the tip density are \cite{bon14}
\begin{eqnarray}
p^+(t,0,y,v,w)=\frac{e^{-|\mathbf{v}-\mathbf{v}_0|^2}}{\int_0^{\infty}\!\int_{-\infty}^{\infty} |\mathbf{v}'|\, e^{-|\mathbf{v}'-\mathbf{v}_0|^2}dv'\,dw'} \nonumber \\
\!\times\! \!\left[j_0(t,y)\! -\!\! \int_{-\infty}^0\!\int_{-\infty}^{\infty}\!\!
|\mathbf{v}'|\, p^-(t,0,y,v',w')d v' dw'\!\right]\!\!, \label{ap3} \\
p^-(t,1,y,v,w)=\frac{e^{-|\mathbf{v}-\mathbf{v}_0|^2}}{\int_{-\infty}^0\!\int_{-\infty}^{\infty}
e^{-|\mathbf{v}'-\mathbf{v}_0|^2}dv'\,dw'} \nonumber\\
\!\times\!  \!\left[\tilde{p}(t,1,y)\! -\!\!\int_0^{\infty}\!\!\int_{-\infty}^{\infty}\! p^+(t,1,y,v',w')dv' dw'\!\right]\!\!, \label{ap4}\\
p(t,\mathbf{x},\mathbf{v})\to 0 \mbox{ as } |\mathbf{v}|\to \infty,\quad\label{ap5}
\end{eqnarray}
where $p^+=p$ for $v>0$ and $p^-=p$ for $v<0$, $\mathbf{v}=(v,w)$. At $x=0$, $j(t,\mathbf{x})$ given by Eq.~\eqref{eq13} is 
\begin{equation}
j_0(t,y)= \alpha(C(t,0,y))\, p(t,0,y,v_0,w_0), \label{ap6}
\end{equation}
for the vector velocity $\mathbf{v}_0=(v_0,w_0)$, with $|\mathbf{v}_0|=1$. The deterministic description including boundary conditions can be proved to have a solution \cite{car16,CDN17}. A convergent numerical scheme to solve the initial boundary value problem corresponding to the deterministic description is studied in Ref.~\cite{bon18}.

\setcounter{equation}{0}
\renewcommand{\theequation}{B.\arabic{equation}}
\section{Collective coordinates for a soliton far from primary vessel and tumor}\label{app3}
To obtain Eqs.~\eqref{eq38}-\eqref{eq40}, we need to substitute the soliton Eq.~\eqref{eq29} into Eq.~\eqref{eq19}. According to Eq.~\eqref{eq29}, the soliton is a function
\begin{equation}
P_s=P_s\!\left(\xi;K,c,\overline{\mu(C)},\overline{F_\xi\!\left(C,\frac{\partial C}{\partial\xi}\right)}\!\right)\!, \label{a1}
\end{equation}
in which we have distinguished the fast coordinate $\xi$ in Eq.~\eqref{eq29} from the slowly varying coordinate $\xi=\psi$ resulting from averages of the TAF density. Assuming that $\mu(C)$ is approximately constant, we have the expressions in Eqs.~\eqref{eq33}, \eqref{eq34} and 
\begin{eqnarray}
\frac{\partial P_s}{\partial t}=\frac{\partial P_s}{\partial K}\dot{K}\!+\!\frac{\partial P_s}{\partial c}\dot{c}\!+\!\frac{\partial P_s}{\partial\overline{F_\xi}} \frac{\partial\overline{F_\xi}}{\partial t}, \label{a2}
\end{eqnarray}
\begin{eqnarray}
\overline{F_\xi}=\frac{\delta}{\beta}\overline{\frac{\hat{\mathbf{V}}\cdot\nabla_x C}{(1+\Gamma_1C)^q}}=\frac{\delta\,\overline{\hat{\mathbf{V}}\cdot\nabla_x(1+\Gamma_1C)^{1-q}}}{\beta(1-q)\Gamma_1} , \label{a3}
\end{eqnarray}

\begin{eqnarray}
&&\frac{\partial\overline{F_\xi}}{\partial t}=\dot{\phi}\overline{F_\eta}+ \frac{\delta/\beta}{(1-q)\Gamma_1} \frac{\partial}{\partial\xi}\overline{\frac{\partial}{\partial t}(1+\Gamma_1C)^{1-q}}\nonumber\\
&&\quad
=\frac{\delta}{\beta}\frac{\partial}{\partial\xi}\!\!\left[\overline{\frac{\frac{\partial C}{\partial t}}{(1\!+\!\Gamma_1C)^q}}\right]\!+\dot{\phi}\overline{F_\eta},\label{a4}
\end{eqnarray}
where we have used $\partial\hat{\mathbf{V}}/\partial t=\dot{\phi}\hat{\mathbf{V}}^\perp$. Setting $\partial C/\partial t=0$ on the right-hand side of Eq.~\eqref{a4}, we obtain
\begin{eqnarray}
\frac{\partial\overline{F_\xi}}{\partial t}=\dot{\phi}\,\overline{F_\eta}. \label{a5}
\end{eqnarray}

Note now that $P_s$ in Eq.~\eqref{eq29} is a function of the ratios $\overline{F_\xi}/c$ and $\xi/c$. Then we have 
\begin{eqnarray}
\frac{\partial P_s}{\partial c}= - \frac{\xi}{c} \frac{\partial P_s}{\partial\xi}- \frac{\overline{F_\xi}}{c}\frac{\partial P_s}{\partial\overline{F_\xi}}\Longrightarrow
\frac{\partial P_s}{\partial\overline{F_\xi}}=- \frac{1}{\overline{F_\xi}} \!\left(\xi\frac{\partial P_s}{\partial\xi}+c\frac{\partial P_s}{\partial c}\right)\!.\label{a6}
\end{eqnarray}

The soliton of Eq.~\eqref{eq29} satisfies
\begin{eqnarray}
(\overline{F_\xi}-c)\frac{\partial P_s}{\partial\xi}=\overline{\mu} P_s-g P_s\int_0^t P_s(\xi(s))\, ds, \label{a7}
\end{eqnarray}
in which $\overline{F_\xi}$ and $\overline{\mu}$ vary slowly with $\xi$ and $t$. Integration by parts shows that  
\begin{eqnarray}
\int_{-\infty}^\infty \eta\,\frac{e^{-\frac{\eta^2}{\sigma^2}}}{\sqrt{\pi}\sigma}\Psi(\eta)d\eta=\frac{\sigma^2}{2}\int_{-\infty}^\infty\frac{e^{-\frac{\eta^2}{\sigma^2}}}{\sqrt{\pi}\sigma}\, \frac{\partial\Psi}{\partial\eta}\, d\eta. \label{a8}
\end{eqnarray}
Using Eq.~\eqref{a8}, Eq.~\eqref{eq19} becomes
\begin{eqnarray}
\frac{\partial\tilde{p}}{\partial t}\!&+&\!\frac{\partial}{\partial\xi}\!\left((F_\xi-c)\tilde{p}+\frac{\sigma^2}{2}\dot{\phi}\frac{\partial\tilde{p}}{\partial\eta}-\frac{1}{2\beta}\frac{\partial\tilde{p}}{\partial\xi}\!\right)\!+\frac{\partial}{\partial\eta}\!\left((F_\eta-\xi\dot{\phi})\tilde{p}-\frac{1}{2\beta}\frac{\partial\tilde{p}}{\partial\eta}\!\right)\! \nonumber\\
\!&=&\!\mu\,\tilde{p} 
 -\Gamma\tilde{p}\!\int_0^t\!\tilde{p}(s,\xi(s),\eta(s))\, ds. \label{a9}
\end{eqnarray}
We now integrate this expression with respect to $\tilde{\eta}=\eta/\sigma$ using the Gaussian approximation Eq.~\eqref{eq22} for $\tilde{p}$. As we set $\sigma\to 0$, the result is Eq.~\eqref{eq23}. When we insert Eqs.~\eqref{a2}, \eqref{a5}, \eqref{a6} and \eqref{a7} into Eq.~\eqref{eq23}, we obtain Eqs.~\eqref{eq35}-\eqref{eq36}. For $\dot{\phi}=0$, the CCEs of the soliton, Eqs.~\eqref{eq44}-\eqref{eq45}, are the same as those found in Ref.~\cite{bon16pre} if we replace $g$ instead of $\Gamma$ in the latter reference.

\end{document}